# Unveiling Four Key Factors for Tire Force Control Allocation in 4WID-4WIS Electric Vehicles at Handling Limits

Ao Lu, Runfeng Li, Yunchang Yu, Ziwang Lu and Guangyu Tian

*Abstract*—**The four-wheel independent drive and four-wheel independent steering (4WID-4WIS) configurations enhance control flexibility and dynamic performance potential for more integrated electric vehicles. This paper comprehensively analyzes the impacts of four key factors on tire force control allocation: vertical load estimation, actuator dynamic characteristics, tire force constraints, and wheel steering precision at handling limits. The study demonstrates that precise vertical load estimation enhances lateral force allocation accuracy. Additionally, the self-compensating effect of lateral tire forces minimizes the impact of small deviations in vertical load estimation on tire force control allocation. A novel control allocation method considering actuator dynamics is introduced, effectively improving yaw rate response and reducing tracking errors. Considering tire-road adhesion and actuator rate constraints, an innovative method to calculate the real-time attainable tire force volume is proposed based on the tire slip ratio and slip angle. Feedforward control with bump steer compensation is implemented to improve wheel steering precision and lateral tire force control accuracy. Matlab/Simulink and Carsim co-simulation results emphasize the importance of these key factors' individual impacts and combined effects. This analysis offers valuable insights for developing advanced tire force control allocation strategies in 4WID-4WIS electric vehicles.**

## I. Introduction

The rapid development of electric vehicle (EV) technology has led to a significant transformation in the automotive industry. More integrated electric vehicles, especially those with 4WID-4WIS configurations, have attracted significant attention in electric vehicle control. The advanced configurations enable precise control of each wheel's driving/braking torque and steering angle, offering great potential for enhancing vehicle stability, maneuverability, and energy efficiency [1]. Various studies investigated different aspects of these configurations, such as torque distribution [2], yaw moment control [3], and integrated chassis control [4].

Tire force control allocation plays a critical role in implementing advanced vehicle control strategies, particularly for more integrated electric vehicles with the over-actuated characteristics provided by the 4WID-4WIS configurations. Several researchers explored different control allocation techniques to optimize tire force distribution, including model predictive control [5], convex optimization [6], and fuzzy logic control [7]. These studies have demonstrated the effectiveness of control allocation in improving vehicle dynamics performance, while the influences of certain factors are often overlooked, especially at handling limits.

Under normal driving conditions, tire forces operate far from their limits, and the impacts of control-related factors are relatively minor. In contrast, during extreme conditions, tires may approach or reach the road adhesion limits, necessitating precise consideration of specific factors to maintain control and stability. Neglecting or oversimplifying these factors in such situations can lead to exceeding tire-road adhesion limits, causing significant trajectory tracking errors and increasing the risk of skidding or loss of control [8]. In the entire process of tire force control allocation, accurate vertical load estimation at the observational stage is essential, directly relating to tire force constraints and ground adhesion utilization. At the execution stage, the dynamic characteristics, rate constraints, and execution precision of physical actuators significantly influence tire force implementation. These aspects are critical in control allocation but are frequently understated.

In recent years, some studies have focused on the effects of different vertical load estimation methods on control allocation [9], emphasizing the importance of optimal tire force distribution [10], [11]. Among various approaches, the most widely proposed method for vertical load estimation considers load transfer due to longitudinal and lateral accelerations [12], [13], [14]. However, under extreme operating conditions, marked by drastic changes in vehicle states like larger pitch and roll angles, and considerable load transfer, the impact of vertical load estimation accuracy on control allocation remains an underexplored research area.

Regarding actuator dynamic characteristics, several researchers investigated the impact of motor and steering system response on control allocation [15], exploring ways to enhance energy efficiency. Some works explored the effects of actuator dynamic response parameter variations on control allocation in electric vehicles [16]. Some researchers considered the dynamic characteristics of different actuators and saturation constraints, and proposed a modular real-time allocation control method [17]. Unlike normal conditions, ignoring the actuator's dynamic characteristics can lead to instability or path deviations as the vehicle fails to achieve the desired force promptly under extreme conditions.

Tire force constraints are essential for ensuring that the control allocation results are feasible. Various constraint formulations have been proposed, such as extremum constraints [18], linear octagon constraints [19], and friction circle constraints [20]. The impact of varying degrees of linearization for friction circle constraints on the algorithm's solution was discussed in [21]. Additionally, actuator rate limitations mean that achieving required tire forces within the adhesion limit, especially lateral tire forces, within a control period (typically 10 ms) is challenging due to restricted wheel angle rate by the steering mechanism. The approach of using a rectangle to approximate the tire force domain by calculating the extreme values of longitudinal and lateral forces based on rate constraints, as suggested in [22], oversimplified the issue. It neglected the significant impact of the coupling between longitudinal and lateral forces on control, especially under extreme conditions. Under regular conditions, inaccurate tire force limit estimations may have little influence on vehicle dynamic performance, as tire forces rarely reach constraint boundaries. However, precise tire boundary estimations become crucial in extreme situations since tire forces often operate at constraint limits.

Considering execution precision, drive motor accuracy is typically high [23], and our main focus is on steering angle control precision. The most significant factor is bump steer, resulting from the suspension's kinematic characteristics. The influence of suspension on wheel steering and mitigation strategies through design were discussed in [24], [25]. However, few studies considered ways to reduce the influence of suspension geometry on wheel steering precision through control. During extreme conditions, minor steering angle deviations can significantly impact tire forces, potentially causing them to exceed adhesion limits. As 4WIS configurations allow independent control of each wheel's steering angle, they provide the opportunity for achieving more precise control.

Under normal driving conditions, tire forces remain within their limits. However, under extreme conditions, the impacts of the factors mentioned above become significant, necessitating a more precise and comprehensive control strategy. This study aims to examine the effects of four key factors: vertical load estimation, actuator dynamic characteristics, tire force constraints, and wheel steering precision. We systematically analyze these factors to develop advanced tire force control allocation strategies, enhancing the handling stability of more integrated electric vehicles in extreme conditions. While our approaches focus on 4WID-4WIS vehicles, they also provide valuable insights into other vehicle configurations. The main novelties are as follows:

1) Precise vertical load estimation enhances lateral force allocation accuracy. The self-compensating effect of lateral tire forces minimizes the impact of small deviations in vertical load estimation on tire force control allocation.
2) A new control allocation method is introduced that considers actuator dynamics based on actuators' time constants. This method effectively improves yaw rate response and reduces tracking errors in various scenarios.
3) An innovative approach for estimating tire force constraints is developed, considering tire-road adhesion and actuator constraints. The proposed convex polygonal method can calculate the real-time attainable force volume using tire slip ratio and slip angle, demonstrating its reliability through simulation results.
4) A novel feedforward control method based on suspension kinematic characteristics is developed. It significantly enhances lateral tire force control accuracy and trajectory tracking performance.

The remainder of this paper is organized as follows. Section II presents the dynamic models, including the vehicle, actuator, tire, and trajectory tracking models. Section III details the hierarchical controller design considering four key factors. Section IV presents the simulation results, discussing each investigated factor and their combined impacts on tire force control allocation. Section V concludes the paper, summarizing the essential findings and their implications.

## II. Dynamic Model

### *A. Vehicle Model*

To thoroughly demonstrate the benefits of 4WID-4WIS vehicles, we construct a double-track vehicle dynamic model with 7-DOF, as depicted in Fig. 1. The longitudinal, lateral, yaw motion and 4-wheel dynamics are presented in (1)-(4):

$$\sum F_x - mgf - \frac{1}{2}C_D A\rho v_x^2 = ma_x = m(\dot{v}_x - \omega_r v_y) \quad (1)$$

$$\sum F_y = ma_y = m(\dot{v}_y + \omega_r v_x) \quad (2)$$

$$\sum M_z = I_z \dot{\omega}_r \quad (3)$$

$$T_{di} - T_{bi} - f_{xi} r = I_{wi}\dot{\omega}_i, i = fl, fr, rl, rr \quad (4)$$

where $m$ represents the vehicle mass, $g$ is the acceleration of gravity, $f$ is the rolling resistance coefficient, $C_D$ is the air drag resistance coefficient, $\rho$ is the air density, and $A$ refers to the vehicle's frontal area. The vehicle's longitudinal and lateral velocities are represented by $v_x$ and $v_y$, respectively, while its longitudinal and lateral accelerations are given by $a_x$ and $a_y$. The yaw rate is denoted by $\omega_r$. $I_z$ indicates the vehicle's yaw inertia around its center of gravity (CG). $I_{wi}$ is the rotational inertia of a single wheel and its drivetrain. $\omega_i$ represents the wheel's rotational speed. $T_{di}$ and $T_{bi}$ denote the drive torque and the braking torque. $r$ is the tire effective rolling radius, and $f_{xi}$ represent the longitudinal force of the front left tire, front right tire, rear left tire, and rear right tire by $i = fl, fr, rl, rr$. Moreover, the total longitudinal force, lateral force, and yaw moment are denoted by $\sum F_x$, $\sum F_y$, and $\sum M_z$ separately, which can be calculated as follows:

$$\sum F = M_f f \quad (5)$$

$$\sum F = \left[\sum F_x, \sum F_y, \sum M_z\right]^T \quad (6)$$

$$M_f = \begin{bmatrix} cos\delta_{fl} & sin\delta_{fl} & -\frac{B}{2}cos\delta_{fl} + asin\delta_{fl} \\ -sin\delta_{fl} & cos\delta_{fl} & acos\delta_{fl} + \frac{B}{2}sin\delta_{fl} \\ cos\delta_{fr} & sin\delta_{fr} & \frac{B}{2}cos\delta_{fr} + asin\delta_{fr} \\ -sin\delta_{fr} & cos\delta_{fr} & acos\delta_{fr} - \frac{B}{2}sin\delta_{fr} \\ cos\delta_{rl} & sin\delta_{rl} & -\frac{B}{2}cos\delta_{rl} - bsin\delta_{rl} \\ -sin\delta_{rl} & cos\delta_{rl} & -bcos\delta_{rl} + \frac{B}{2}sin\delta_{rl} \\ cos\delta_{rr} & sin\delta_{rr} & \frac{B}{2}cos\delta_{rr} - bsin\delta_{rr} \\ -sin\delta_{rr} & cos\delta_{rr} & -bcos\delta_{rr} - \frac{B}{2}sin\delta_{rr} \end{bmatrix}^T \quad (7)$$

$$f = \left[f_{xfl}, f_{yfl}, f_{xfr}, f_{yfr}, f_{xrl}, f_{yrl}, f_{xrr}, f_{yrr}\right]^T \quad (8)$$

where $f_{xi}, f_{yi}, \delta_i$ represent the longitudinal force, lateral force, and wheel steering angle for each tire, respectively. $a$ and $b$ represent the distances from the front and rear axles to the vehicle's CG, while $B$ signifies the wheel track.

TABLE I
TIME CONSTANT OF THE ACTUATOR AND TIRE

| Parameter | Symbol | Value |
|---|---|---|
| Drive system time constant | $\tau_d$ | 0.015s |
| Braking system time constant | $\tau_b$ | 0.06s |
| Steering system time constant | $\tau_s$ | 0.1s |
| Maximum/minimum torque | $T_{i,\max\,(min)}$ | $\pm$1250Nm |
| Maximum/minimum torque rate | $\dot{T}_{i,\max\,(min)}$ | $\pm$500Nm/10ms |
| Maximum/minimum wheel steering angle | $\delta_{i,\max\,(min)}$ | $\pm$35° |
| Maximum/minimum wheel steering angle rate | $\dot{\delta}_{i,\max\,(min)}$ | $\pm$0.5°/10ms |
| Front wheel longitudinal force time constant | $\tau_{ffx}$ | 0.014s |
| Front wheel lateral force time constant | $\tau_{ffy}$ | 0.018s |
| Rear wheel longitudinal force time constant | $\tau_{rfx}$ | 0.02s |
| Rear wheel lateral force time constant | $\tau_{rfy}$ | 0.024s |

### B. Actuator Model

Electric vehicle actuators primarily comprise various electromechanical systems for driving, braking, and steering functions. These systems respond differently to the tire forces exerted during operation.

The driving torque is achieved through the control of the drive motor. Assuming the motor control strategy maintains the direct-axis current at zero, the dynamic response can be approximated as a first-order inertia link during the motor control process [26]. Meanwhile, the physical models of the electronic hydraulic braking system and the independent steering system can be approximated as first-order inertia links [27], [28] through PID control:

$$G(s) = \frac{1}{\tau s+1} \tag{9}$$

where $\tau$ represents the time constant, obtained from real vehicle test platforms [29] and detailed in Table I. The drive system's time constant, measured on a motor test bench, is approximately 15 ms. The braking system's time constant, assessed via its test platform, is about 60 ms. Dynamic characteristic tests of the steering system indicate a time constant of around 100 ms, revealing a notably greater inertial lag in the steering system.

### C. Tire Model

A combined brush model is employed to demonstrate the nonlinear tire behavior:

$$\begin{cases} f_x = \frac{K_s}{\gamma_t}\left(\frac{\kappa}{1+\kappa}\right) f, f_y = \frac{-K_\alpha}{\gamma_t}\left(\frac{\tan\alpha}{1+\kappa}\right) f \\ \gamma_t = \sqrt{K_s^2\left(\frac{\kappa}{1+\kappa}\right)^2 + K_\alpha^2\left(\frac{\tan\alpha}{1+\kappa}\right)^2} \\ f = \begin{cases} \gamma_t - \frac{1}{3\mu f_z}\gamma_t^2 + \frac{1}{27\mu^2 f_z^2}\gamma_t^3, (\gamma_t \le 3\mu f_z) \\ \mu f_z, (\gamma_t > 3\mu f_z) \end{cases} \end{cases} \tag{10}$$

where $K_s$ and $K_\alpha$ are the tire longitudinal stiffness and cornering stiffness, respectively. $\kappa$ is the slip rate of the tire, and $\alpha$ is the tire slip angle. The conditions for tire longitudinal and lateral slip are depicted in Fig. 2 and Fig. 3, respectively.

When considering tire dynamic characteristics, the tire force exhibits relaxation and lag characteristics under dynamic input. This dynamic response can be approximated as a first-order system [30], as shown below:

$$f_{xD} = \frac{1}{1+\tau_x s} f_{xS},\ f_{yD} = \frac{1}{1+\tau_y s} f_{yS} \tag{11}$$

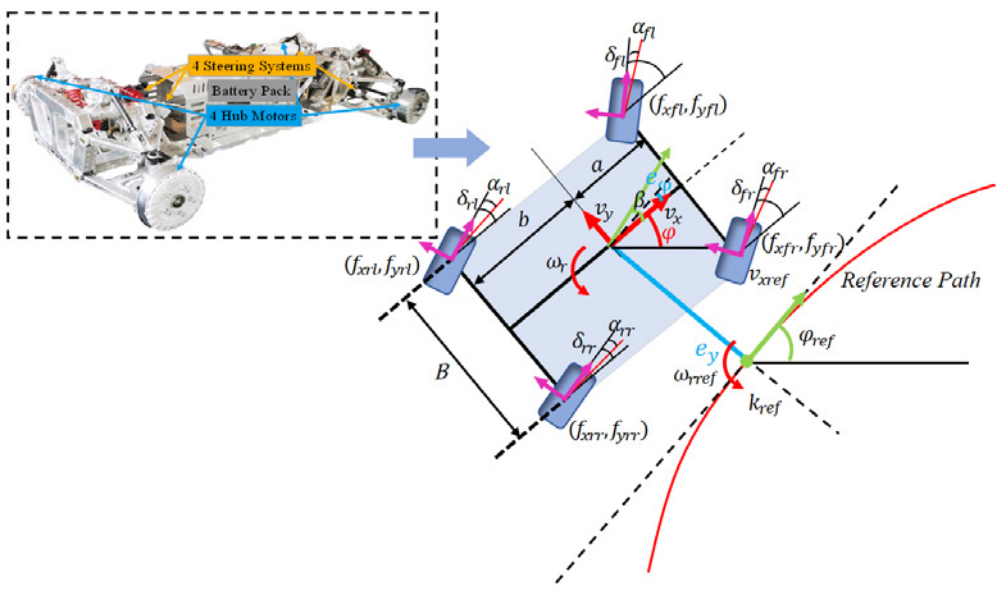

Fig. 1. The vehicle model and the trajectory tracking model.

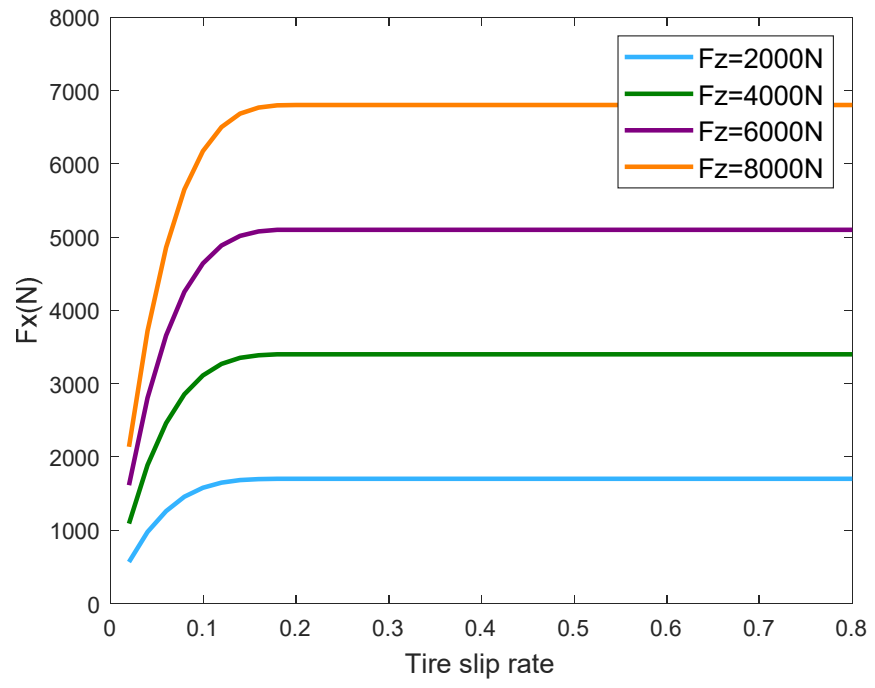

Fig. 2.Tire pure longitudinal slip condition.

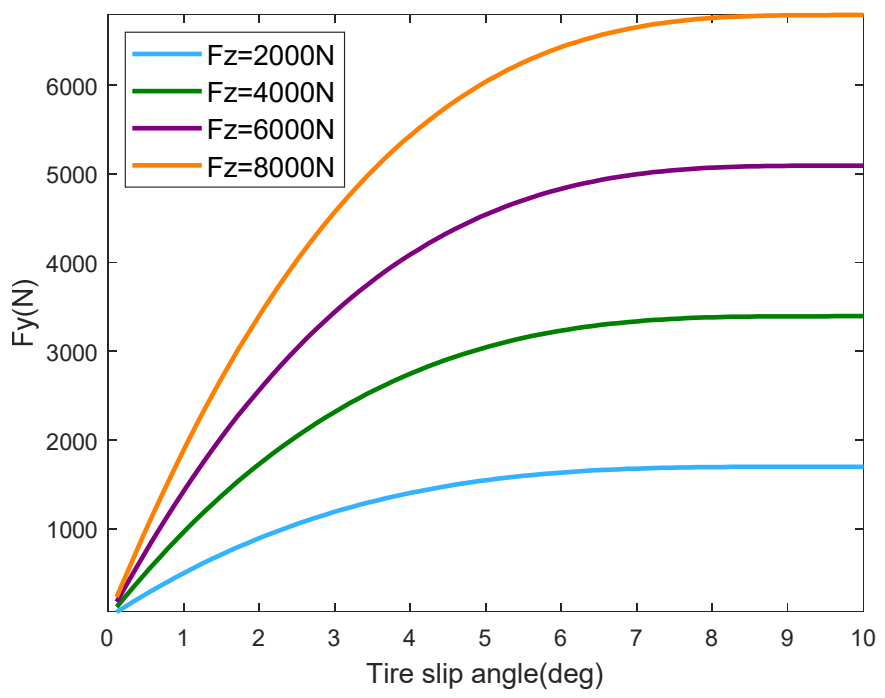

Fig. 3.Tire pure lateral slip condition.

where $f_{xD}, f_{yD}$ and $f_{xS}, f_{yS}$ are the dynamic tire force and the steady-state tire force, respectively.

### D. Trajectory Tracking Model

The trajectory tracking model aims to guide the vehicle to the desired trajectory. We minimize the tracking errors and track the target states accurately by adopting four parameters, ensuring the vehicle maintains the desired trajectory, speed, and orientation, as shown in Fig. 1. These parameters include the longitudinal velocity error $e_{vx}$, the head error $e_\varphi$, the lateral error $e_y$ and the yaw rate error $e_{\omega r}$, which can be calculated by:

$$e_{vx} = v_{xref} - v_x \tag{12}$$

$$e_\varphi = \varphi_{ref} - \varphi \tag{13}$$

$$\dot{e}_y = v_x e_\varphi + v_y \tag{14}$$

$$e_{\omega r} = \omega_r - \omega_{rref} = \omega_r - \frac{v_{xref}}{\cos(e_\varphi)} k_{ref} \tag{15}$$

where $\varphi$ is the yaw angle, and $v_{xref}$, $\varphi_{ref}$, $k_{ref}$ are respectively the reference longitudinal speed, yaw angle and trajectory curvature.

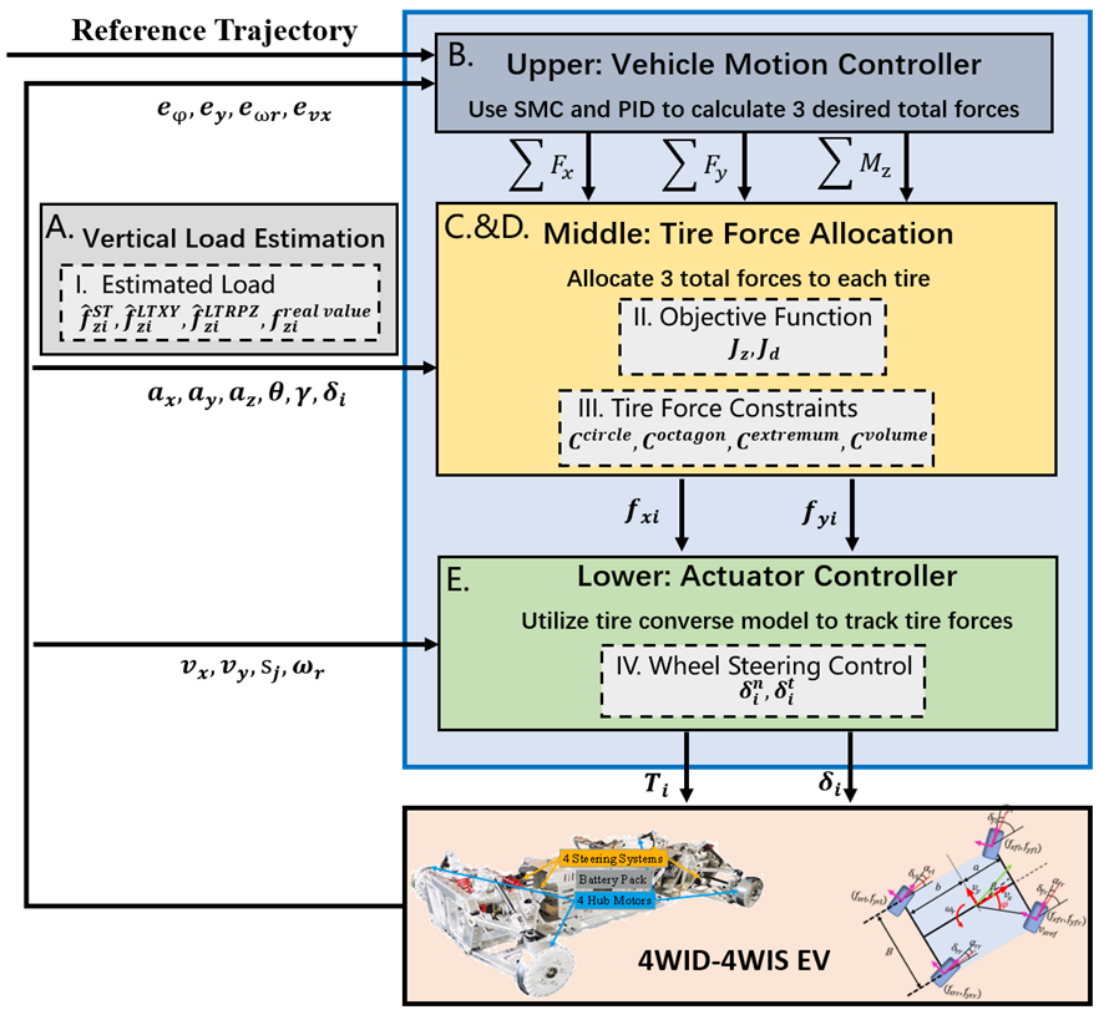

Fig. 4. The hierarchical architecture of tire force control allocation.

## III. Tire Force Control Allocation Design

A hierarchical architecture is designed for distributing desired tire forces and moments to each wheel. The control framework comprises three layers, as illustrated in Fig. 4. The upper-layer motion controller determines the overall force requirements based on trajectory tracking. The middle-layer control allocation distributes tire forces to the lower-layer actuators for control execution.

### A. Vertical Load Estimation

Before carrying out tire force control allocation, accurately estimating the tire's vertical load is crucial, as it is the factor most closely related to control allocation at the observational stage. This estimation influences the determination of tire force and the steering angle execution process, ultimately affecting ground adhesion utilization.

The static load estimation (ST) is the most simplified model, ignoring dynamic load transfer:

$$\hat{f}_{zfl}^{ST} = \hat{f}_{zfr}^{ST} = \frac{mgb}{2L}, \hat{f}_{zrl}^{ST} = \hat{f}_{zrr}^{ST} = \frac{mga}{2L} \tag{16}$$

where $L$ represents the vehicle's wheelbase.

The vertical load estimation with longitudinal and lateral load transfer (LTXY) neglects the pitch and roll effects of sprung mass:

$$\begin{cases} \hat{f}_{zfl}^{LTXY} = \frac{mb}{2L}\left(g - \frac{a_x h}{b} - \frac{2a_y h}{B}\right) \\ \hat{f}_{zfr}^{LTXY} = \frac{mb}{2L}\left(g - \frac{a_x h}{b} + \frac{2a_y h}{B}\right) \\ \hat{f}_{zrl}^{LTXY} = \frac{ma}{2L}\left(g + \frac{a_x h}{a} - \frac{2a_y h}{B}\right) \\ \hat{f}_{zrr}^{LTXY} = \frac{ma}{2L}\left(g + \frac{a_x h}{a} + \frac{2a_y h}{B}\right) \end{cases} \tag{17}$$

This method is widely used, only requiring longitudinal and lateral acceleration sensors, which are generally available in most vehicles.

Considering the vehicle body roll, pitch and vertical acceleration, as illustrated in Fig. 5, the load transfer (LTRPZ) estimation is better than the two methods mentioned above. Second-order terms are neglected during the derivation process, resulting in the following expressions:

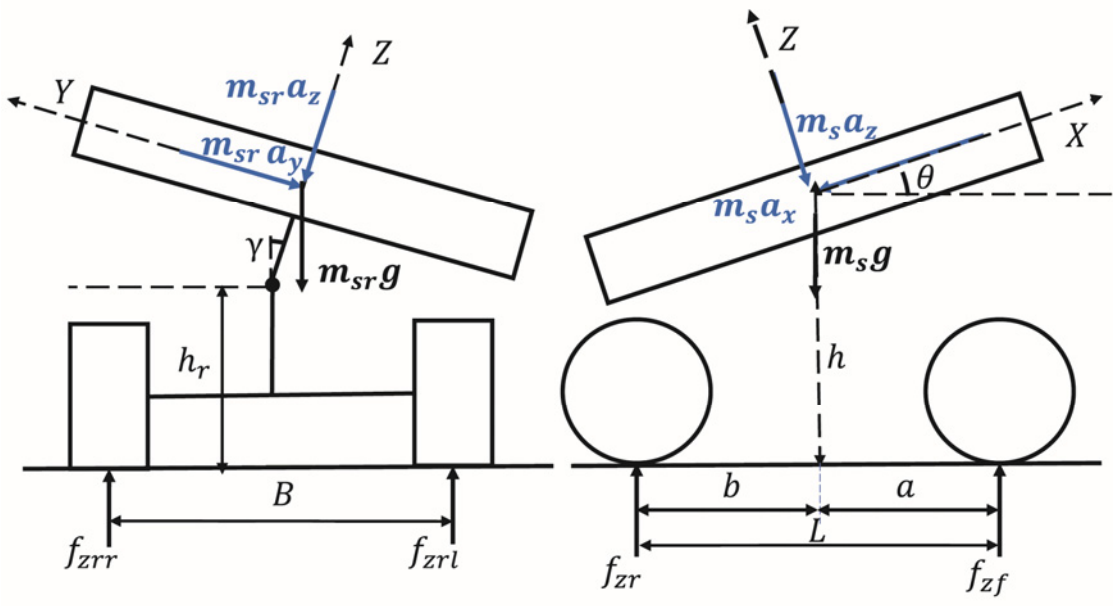

Fig. 5. Vehicle roll and pitch model.

$$\begin{cases} \hat{f}_{zfl}^{LTRPZ} = \frac{mbg + m_s\cos\theta(ba_z - a_x h)}{2L} - \frac{\frac{m_s a_y b h_r}{L} + m_u a_y r + K_1\gamma}{B} \\ \hat{f}_{zfr}^{LTRPZ} = \frac{mbg + m_s\cos\theta(ba_z - a_x h)}{2L} + \frac{\frac{m_s a_y b h_r}{L} + m_u a_y r + K_1\gamma}{B} \\ \hat{f}_{zrl}^{LTRPZ} = \frac{mag + m_s\cos\theta(aa_z + a_x h)}{2L} - \frac{\frac{m_s a_y a h_r}{L} + m_u a_y r + K_2\gamma}{B} \\ \hat{f}_{zrr}^{LTRPZ} = \frac{mag + m_s\cos\theta(aa_z + a_x h)}{2L} + \frac{\frac{m_s a_y a h_r}{L} + m_u a_y r + K_2\gamma}{B} \end{cases} \tag{18}$$

where $m_s$ and $m_u$ represent the sprung mass and unsprung mass of the vehicle, respectively. $a_z$ is the vertical acceleration. $\theta$ is the pitch angle of the vehicle. $\gamma$ is the roll angle of the vehicle. $h_r$ is the roll center height. $K_1$ and $K_2$ are front axle roll stiffness and rear axle roll stiffness.

The actual load serves as a reference for comparison, reflecting the impact of vertical load estimation on control allocation and the execution of tire forces.

### B. Vehicle Motion Control

At the upper level, the vehicle motion controller employs two separate PI controllers to calculate the required total longitudinal force and yaw moment based on the yaw rate error and the longitudinal velocity error. A sliding mode control (SMC) is employed for its robustness against uncertainties and variabilities [31]. It is designed [32] to minimize lateral tracking error and achieve the resultant lateral force.

### C. Tire Force Allocation

At the middle level, the total demanded forces and moment are allocated to each tire based on control allocation. We propose a new control allocation method that considers actuator dynamic characteristics while building upon the commonly used objective of enhancing tire adhesion utilization margin.

The primary optimization objective is to satisfy the total control demand as much as possible, considering road adhesion and actuator constraints. To avoid infinitely many solutions that satisfy the primary objective, a secondary optimization objective of minimizing the total tire adhesion utilization is added. The optimization objective $J_z$ can be expressed as:

$$J_z = \min\left(\left\|k_\gamma W_f f\right\|_2^2 + \left\|\sum F - M_f f\right\|_2^2\right) \tag{19}$$

$$W_f = diag\left(\frac{1}{\mu\hat{f}_{zfl}^j}, \frac{1}{\mu\hat{f}_{zfl}^j}, \frac{1}{\mu\hat{f}_{zfr}^j}, \frac{1}{\mu\hat{f}_{zfr}^j}, \frac{1}{\mu\hat{f}_{zrl}^j}, \frac{1}{\mu\hat{f}_{zrl}^j}, \frac{1}{\mu\hat{f}_{zrr}^j}, \frac{1}{\mu\hat{f}_{zrr}^j}\right) \tag{20}$$

where $\mu$ represents the road surface adhesion coefficient, and $k_\gamma$ is a weighting factor. $\hat{f}_{zi}^j, j = ST, LTXY, LTRPZ$ represent the estimated vertical force of each wheel. The observation of vertical load directly affects the weight matrix $W_f$, which influences the distribution ratio among different wheels.

Actuator dynamics significantly affect tire force distribution in control allocation. Tires also exhibit lag under dynamic input. These factors can be modeled as first-order inertial links, represented as:

$$u_o(t+\Delta T) = \mathrm{e}^{-\frac{\Delta T}{\tau}} u_o(t) + \left(1 - \mathrm{e}^{-\frac{\Delta T}{\tau}}\right) u_i(t) \tag{21}$$

where $u_o$ is the output and $u_i$ is the input. $\tau$ represents the time constants of each actuator. $\Delta T$ represents the control period.

Therefore, we propose a tire force control allocation method based on actuator dynamic characteristics. It is assumed that the increment of tire force output is proportional to the bandwidth of each actuator. Specifically, we add a penalty term related to actuator speed in the optimization objective, allowing faster-responding actuators to be allocated more when the global control demand changes rapidly. The new optimization objective $J_d$ can be expressed as:

$$J_d = \min \left( \left\| k_d W_{df} (f - f_{k-1}) \right\|_2^2 + J_z \right) \tag{22}$$

$$W_{df} = diag\left(\frac{1}{\omega_{xfl}} \frac{1}{\omega_{yfl}} \frac{1}{\omega_{xfr}} \frac{1}{\omega_{yfr}} \frac{1}{\omega_{xrl}} \frac{1}{\omega_{yrl}} \frac{1}{\omega_{xrr}} \frac{1}{\omega_{yrr}}\right) \tag{23}$$

where $k_d$ is a weighting factor, and $f_{k-1}$ is the tire force allocation at the previous time step. $\omega_{xi}$ and $\omega_{yi}$ represent the bandwidth of the tire forces on each side, which can be approximated by merging the small parameters of the actuator time constants [22].

Additionally, we perform compensation and correction on the time-domain response of the first-order actuator model:

$$u_i(t) = \frac{1}{1 - \mathrm{e}^{-\frac{\Delta T}{\tau}}} \left(u_{cmd}(t) - u_o(t)\right) + u_o(t) \tag{24}$$

where $u_{cmd}$ denotes the desired actuator output given by the control allocation.

### *D. Tire Attainable Force Volume Calculation*

The attainable force volume of tires [33] can be calculated, factoring in road adhesion and actuator constraints. It serves as a crucial reference for determining the required tire force within limits during control allocation, impacting actual control accuracy. Friction circle constraints describe the relationship between the longitudinal and lateral forces at the tire-road interface:

$$\sqrt{f_{xi}^2 + f_{yi}^2} \le \mu \hat{f}_{zi}^{j} \tag{25}$$

Friction circle constraints transform control allocation into a classic quadratically constrained quadratic programming (QCQP) problem, potentially raising computational effort due to increased algorithm execution time. The distribution algorithm employs sequential quadratic programming (SQP) to compute required tire forces [34]. For optimization simplification and computational efficiency enhancement [21], the friction circle is frequently approximated by a linear octagon. This turns the optimization challenge into a classic quadratic programming (QP) problem. The active-set algorithm is employed to solve the problem because of its fast computational efficiency. The extremum constraints are the simplest physical limitation of road adhesion on tire force, ignoring the tire combined slip in the vehicle system:

$$|f_{xi}| \le \mu \hat{f}_{zi}^{j}, |f_{yi}| \le \mu \hat{f}_{zi}^{j}. \tag{26}$$

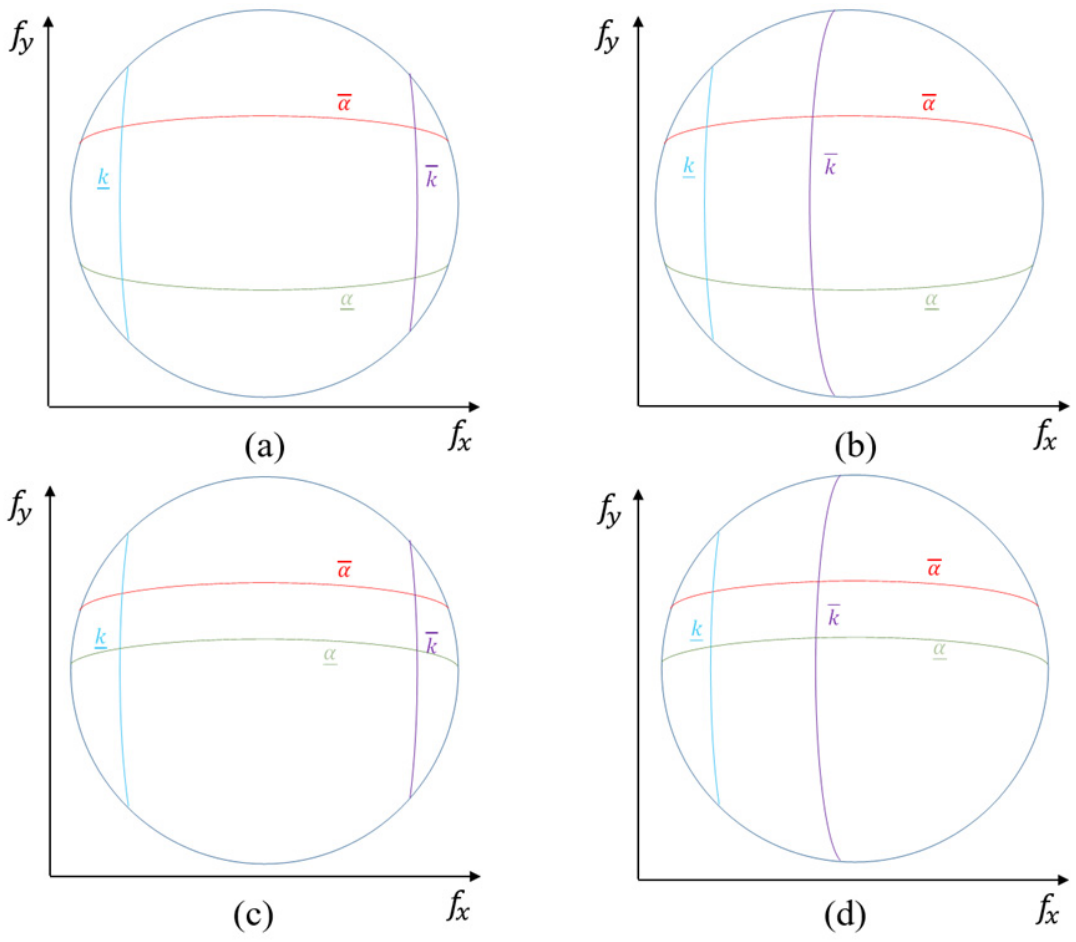

Fig. 6. The tire attainable tire force volumes under different limits of tire slip ratio and tire slip angle. (a) $\underline{\alpha_i} \le 0 \le \overline{\alpha}_i$, $\underline{\kappa_i} \le 0 \le \overline{\kappa}_i$; (b) $\underline{\alpha_i} \le 0 \le \overline{\alpha}_i$, $\underline{\kappa_i} \le \overline{\kappa}_i \le 0$; (c) $0 \le \underline{\alpha_i} \le \overline{\alpha}_i$, $\underline{\kappa_i} \le 0 \le \overline{\kappa}_i$; (d) $0 \le \underline{\alpha_i} \le \overline{\alpha}_i$, $\underline{\kappa_i} \le \overline{\kappa}_i \le 0$.

Besides tire-to-road adhesion constraints, the consideration of actuator constraints, especially rate constraints, is crucial but often overlooked. In this section, a new method for real-time calculation of the attainable tire force volume is proposed. Given the constraints on torque and wheel steering angle, both in magnitude and rate as illustrated in Table I, the attainable tire force volume continuously changes due to limitations in driving/braking and steering systems.

Based on the constraints of torque in terms of magnitude and rate, the torque constraints for the wheels at the next moment can be calculated in real time as:

$$\underline{T_i} \le T_i \le \overline{T}_i \tag{27}$$

$$\underline{T_i} = \max\left(T_{i,min},\ T_i(k) + \dot{T}_{i,min} \Delta T\right) \tag{28}$$

$$\overline{T}_i = \min\left(T_{i,max},\ T_i(k) + \dot{T}_{i,max} \Delta T\right) \tag{29}$$

Using the wheel dynamics equations (4), the range of wheel speeds can be determined. Assuming the current vehicle speed remains constant, the slip ratio $\kappa_i$ constraints for the next moment are as follows:

$$\kappa_i = \frac{\omega_i r - v_x}{|v_x|} \tag{30}$$

$$\underline{\kappa_i} \le \kappa_i \le \overline{\kappa}_i \tag{31}$$

Based on the constraints of steering angle in terms of magnitude and rate, the steering angle constraints for the wheels at the next moment can be calculated in real time as follows:

$$\underline{\delta_i} \le \delta_i \le \overline{\delta}_i \tag{32}$$

$$\underline{\delta_i} = \max\left(\delta_{i,min},\ \delta_i(k) + \dot{\delta}_{i,min} \Delta T\right) \tag{33}$$

$$\overline{\delta}_i = \min\left(\delta_{i,max},\ \delta_i(k) + \dot{\delta}_{i,max} \Delta T\right) \tag{34}$$

The tire slip angle $\alpha_i$ constraints can be derived from the formula for calculating the slip angle as follows:

$$\begin{aligned} \alpha_{fl} &= \tan^{-1}\left(\frac{v_y + a\omega_r}{v_x - \frac{B}{2}\omega_r}\right) - \delta_{fl}, \alpha_{fr} = \tan^{-1}\left(\frac{v_y + a\omega_r}{v_x + \frac{B}{2}\omega_r}\right) - \delta_{fr} \\ \alpha_{rl} &= \tan^{-1}\left(\frac{v_y - b\omega_r}{v_x - \frac{B}{2}\omega_r}\right) - \delta_{rl}, \alpha_{rr} = \tan^{-1}\left(\frac{v_y - b\omega_r}{v_x + \frac{B}{2}\omega_r}\right) - \delta_{rr} \end{aligned} \tag{35}$$

$$\underline{\alpha_i} \le \alpha_i \le \overline{\alpha}_i \tag{36}$$

The attainable domain of tire forces can be ascertained using $\underline{k}_i$, $\overline{k}_i$, $\underline{\alpha}_i$ and $\overline{\alpha}_i$, based on the joint slip brush model, as introduced in (10). Considering the symmetry in driving and braking as well as left and right turning, four major categories of attainable tire force volumes are identified, as depicted in Fig. 6. Notably, only category (a) forms a convex set, while the others are non-convex, presenting mathematical complexity in description. The non-convex nature of these constraints makes solving the associated optimization problem NP-hard. To enhance computational efficiency, a convex polygonal approximation method is employed to approximate the attainable force volume, as shown in Fig. 7.

Using the convex polygonal method to approximate the attainable domain of tire forces, the specific calculation rules for each vertex are detailed in Table II. The vertices $A_1$ to $A_8$ are arranged in alphabetical order to form a polygon for real-time approximation of the tire force domain, as illustrated in Fig. 7. Notably, in categories (b), (c) and (d), when the minimum slip ratio or tire slip angle exceeds 0, or the maximum value is below 0, vertices may overlap, and the actual region becomes a non-convex polygon. To improve computational efficiency, the attainable domain of tire forces is represented by a convex polygon formed by vertices, which imposes real-time linear constraints on tire forces.

### E. Actuator Control

At the lower level, the calculated tire forces are achieved by each wheel's torque and steering angle. The wheel driving/braking torque is determined based on each tire's longitudinal force, as (4) shows. The normal steering angles $\delta_i^n$ of four wheels can be demonstrated from (35). $\alpha_i$ is obtained through the tire inverse model, for which a look-up table is created to address the challenge of tire force nonlinearity. Depending on the lateral tire force and vertical tire load, the required tire slip angle is determined.

During the wheel steering process, bump steer is the primary factor affecting wheel steering angle precision. It is mainly influenced by toe angle changes due to suspension kinematic characteristics, which significantly impacts lateral tire force control accuracy. To compensate for bump steer, we propose a feedforward controller based on suspension displacement data from suspension sensors, depicted in Fig. 8. Taking the front left wheel as an example, the adjusted steering angle $\delta_{fl}^t$ can be derived as:

$$\delta_{fl}^t = \tan^{-1}\left(\frac{v_y + a\omega_r}{v_x - \frac{B}{2}\omega_r}\right) - \alpha_{fl} + \delta_{sfl} \tag{37}$$

where $\delta_{sfl}$ is the bump steer angle for the front left tire resulting from suspension displacement $s_{jfl}$.

## IV. Simulation and Results

Based on the discussion in Section III, we investigate the effects of three types of estimated vertical load (ST, LTXY, LTRPZ), two optimization objective functions (control considering actuator dynamics, control neglecting actuator dynamics), four tire force constraints (friction circle, linear octagon, extremum, attainable force volume), and two kinds of wheel steering methods (feedforward compensation control, normal steer) on control allocation in this section. We perform joint simulations using Matlab/Simulink and Carsim, focusing on a 4WID-4WIS EV with parameters listed in Table III. We utilize a step yaw rate input maneuver (80 km/h, targeting 22 deg/s yaw rate at 1 s) to analyze the steady-state steering characteristics under extreme conditions. Additionally, we conduct double lane change (80 km/h) and slalom [35] (80 km/h, 30 m inter-cone distance) maneuvers to analyze the transient steering characteristics. The road surface adhesion coefficient is consistently set to 1.

TABLE II
Rules for Calculating Vertices in Convex Polygonal Approximation Method

| Vertex | Calculation Rules |
|---|---|
| $A_1$ | $\alpha = \overline{\alpha}_i, \kappa = 0, if\ \underline{\kappa}_i \le 0 \le \overline{\kappa}_i$<br>$\alpha = \overline{\alpha}_i, \kappa = \underline{\kappa}_i, if\ \underline{\kappa}_i > 0$<br>$\alpha = \overline{\alpha}_i, \kappa = \overline{\kappa}_i, if\ \overline{\kappa}_i < 0$ |
| $A_2$ | $\alpha = \overline{\alpha}_i, \kappa = \overline{\kappa}_i, if\ \gamma_t(\overline{\alpha}_i, \overline{\kappa}_i) \le 3\mu f_z$<br>$\alpha = \overline{\alpha}_i, \kappa = \overline{\kappa}_{inew}(\gamma_t(\overline{\alpha}_i, \overline{\kappa}_{inew}) = 3\mu f_z), if\ \gamma_t(\overline{\alpha}_i, \overline{\kappa}_i) > 3\mu f_z$ |
| $A_3$ | $\alpha = 0, \kappa = \overline{\kappa}_i, if\ \underline{\alpha}_i \le 0 \le \overline{\alpha}_i$<br>$\alpha = \underline{\alpha}_i, \kappa = \overline{\kappa}_i, if\ \underline{\alpha}_i > 0$<br>$\alpha = \overline{\alpha}_i, \kappa = \overline{\kappa}_i, if\ \overline{\alpha}_i < 0$ |
| $A_4$ | $\alpha = \underline{\alpha}_i, \kappa = \overline{\kappa}_i, if\ \gamma_t(\underline{\alpha}_i, \overline{\kappa}_i) \le 3\mu f_z$<br>$\alpha = \underline{\alpha}_i, \kappa = \overline{\kappa}_{inew}(\gamma_t(\underline{\alpha}_i, \overline{\kappa}_{inew}) = 3\mu f_z), if\ \gamma_t(\underline{\alpha}_i, \overline{\kappa}_i) > 3\mu f_z$ |
| $A_5$ | $\alpha = \underline{\alpha}_i, \kappa = 0, if\ \underline{\kappa}_i \le 0 \le \overline{\kappa}_i$<br>$\alpha = \underline{\alpha}_i, \kappa = \underline{\kappa}_i, if\ \underline{\kappa}_i > 0$<br>$\alpha = \underline{\alpha}_i, \kappa = \overline{\kappa}_i, if\ \overline{\kappa}_i < 0$ |
| $A_6$ | $\alpha = \underline{\alpha}_i, \kappa = \underline{\kappa}_i, if\ \gamma_t(\underline{\alpha}_i, \underline{\kappa}_i) \le 3\mu f_z$<br>$\alpha = \underline{\alpha}_i, \kappa = \underline{\kappa}_{inew}(\gamma_t(\underline{\alpha}_i, \underline{\kappa}_{inew}) = 3\mu f_z), if\ \gamma_t(\underline{\alpha}_i, \underline{\kappa}_i) > 3\mu f_z$ |
| $A_7$ | $\alpha = 0, \kappa = \underline{\kappa}_i, if\ \underline{\alpha}_i \le 0 \le \overline{\alpha}_i$<br>$\alpha = \underline{\alpha}_i, \kappa = \underline{\kappa}_i, if\ \underline{\alpha}_i > 0$<br>$\alpha = \overline{\alpha}_i, \kappa = \underline{\kappa}_i, if\ \overline{\alpha}_i < 0$ |
| $A_8$ | $\alpha = \overline{\alpha}_i, \kappa = \underline{\kappa}_i, if\ \gamma_t(\overline{\alpha}_i, \underline{\kappa}_i) \le 3\mu f_z$<br>$\alpha = \overline{\alpha}_i, \kappa = \underline{\kappa}_{inew}(\gamma_t(\overline{\alpha}_i, \underline{\kappa}_{inew}) = 3\mu f_z), if\ \gamma_t(\overline{\alpha}_i, \underline{\kappa}_i) > 3\mu f_z$ |

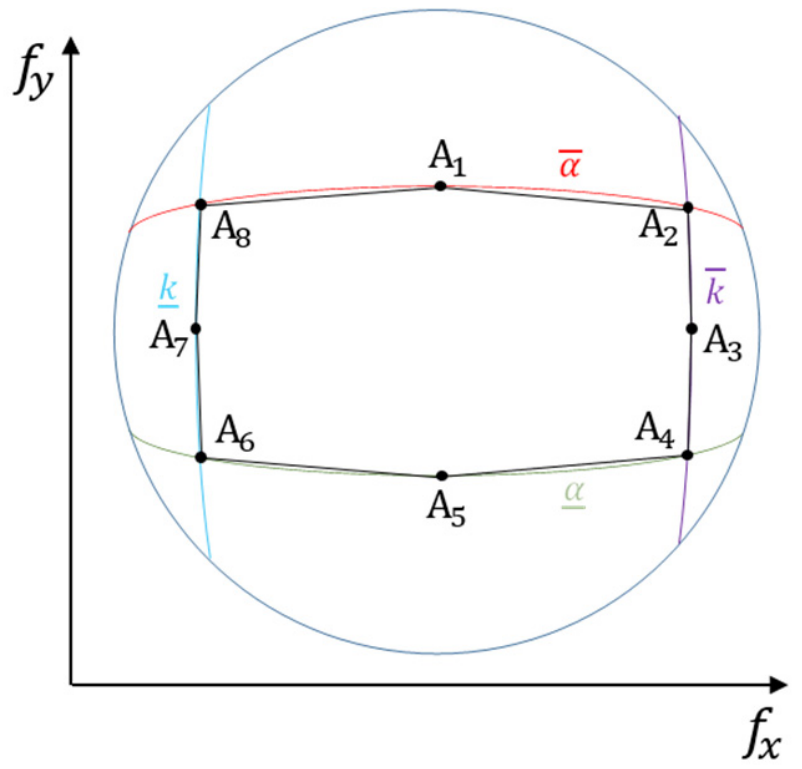

Fig. 7. Convex polygonal method for approximating the attainable domain of tire forces (illustrated with category (a)).

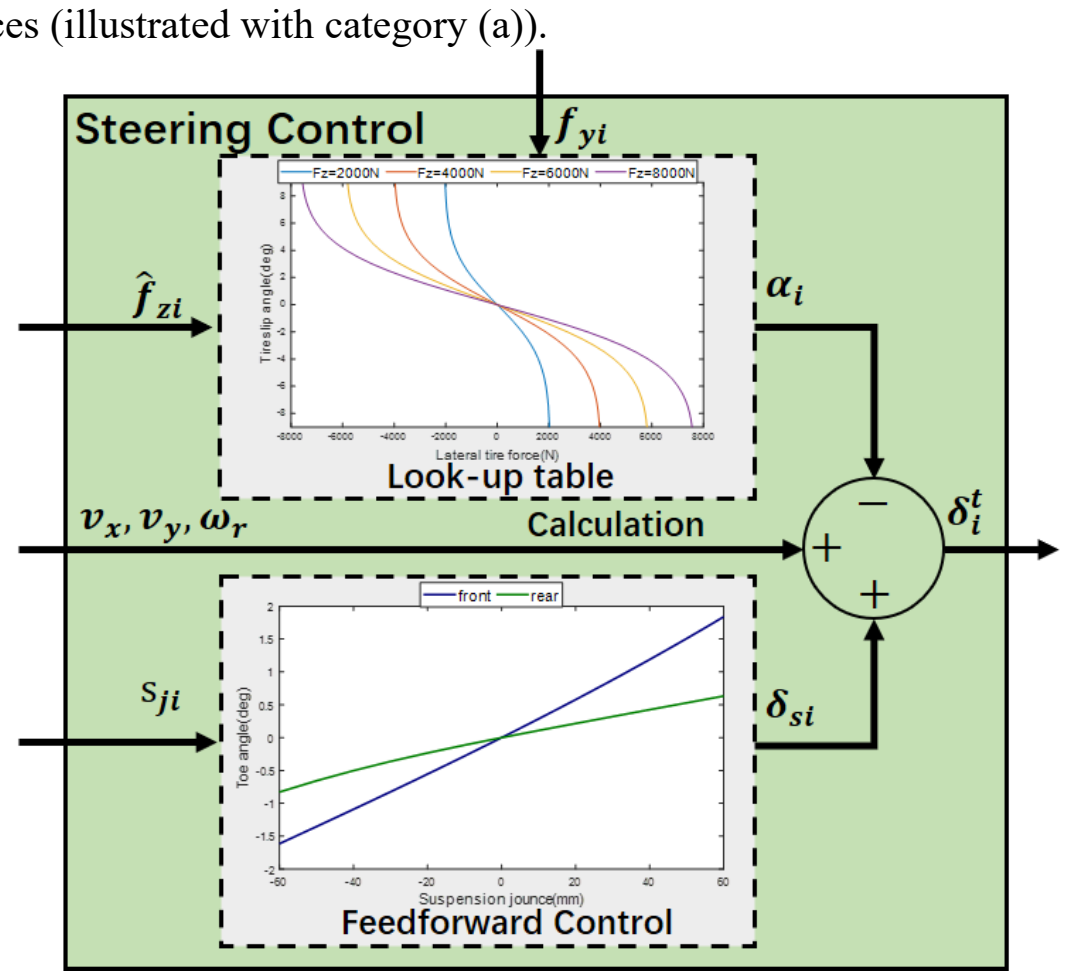

Fig. 8. Wheel steering angle control architecture at the execution stage.

TABLE III
PARAMETERS OF THE 4WID-4WIS EV MODEL

| Parameter | Symbol | Value |
|---|---|---|
| Sprung mass | $m_s$ | 1110kg |
| Unsprung mass (both sides) | $m_u$ | 60kg |
| Vehicle yaw inertia | $I_z$ | 1343.1kg·m$^{-2}$ |
| CG height | $h$ | 0.54m |
| Front wheelbase | $a$ | 1.06m |
| Rear wheelbase | $b$ | 1.54m |
| Roll center height | $h_r$ | 0.5 |
| Tire effective rolling radius | $r$ | 0.298m |
| Front axle roll stiffness | $K_1$ | 60N·m/° |
| Rear axle roll stiffness | $K_2$ | 150 N·m/° |
| Wheel track | $B$ | 1.48m |

In the comparison process, uniform controller parameters are utilized. The analysis uses a baseline model featuring wheel steering angle control with feedforward compensation, based on the real value of vertical load, and without actuator dynamic characteristics or rate constraints, for conducting a comparative analysis of single-factor influence.

### A. *Vertical Load Estimation*

The comparative results of three different vertical load estimation methods are illustrated in Table IV. The method which takes into account the roll angle, pitch angle, and vertical acceleration, outperforms the other two methods in all scenarios. Notably, in the slalom scenario where the vehicle turns most frequently, the estimation error of LTRPZ decreases by 4% in contrast with the method that only considers load change caused by acceleration. When a vehicle operates under extreme maneuvers, rapid acceleration, and emergency turns often occur, inevitably causing roll and pitch angles, which will affect the estimation accuracy.

As shown in Fig. 8, lateral actuator control requires the look-up table with the nonlinear model. If vertical load estimation error exists, the tire slip angle will deviate from the ideal value, leading to lateral force deviation and influencing vehicle control performance. Fig. 9 depicts that under the step yaw rate condition, LTRPZ outperforms other load estimation methods, closely matching the response with the real value. The accurate vertical load estimation can also suppress the vehicle sideslip angle due to improved lateral force allocation accuracy. Furthermore, precise vertical load estimation contributes to improved tire force execution, enabling quicker attainment of the vehicle's ideal state and consequently reducing response time.

The influence of vertical load estimation on control allocation can be viewed from another perspective. As illustrated in Fig. 4 and (20), the tire vertical force estimation results determine the weighting function of each tire. Abe discussed the impact of different weight coefficients on stability control [36]. He proposed that when the distribution weight of the front axle is greater than that of the rear axle, the vehicle response is more sensitive; when the distribution weight of the rear axle is greater than that of the front axle, the vehicle tends to be more stable. This is intuitive as the more extensive weight results in the smaller tire force. Similarly, biased vertical force estimation will allocate more tire force to the tire with overestimated vertical force. However, when the scenarios demand mostly tire lateral forces, like the double lane change (DLC) and slalom defined in this article, the tracking error differences remain minimal as Fig. 10 illustrates. It is due to the self-compensating effect of the control allocation layer and actuator control layer using the look-up table method when determining wheel steering angles, which will be analyzed in detail in the following paragraph.

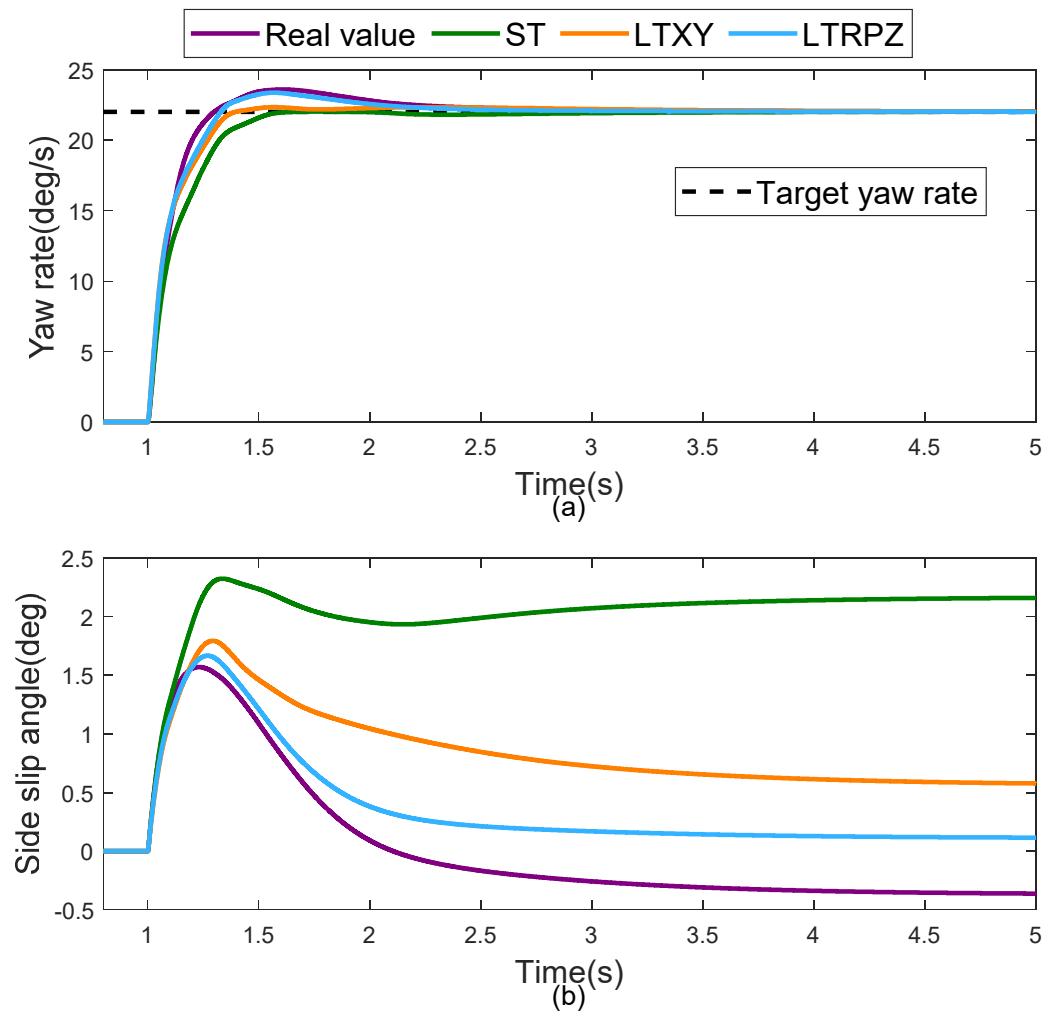

Fig. 9. Transient response results with vertical load estimation in step yaw rate input. (a) Yaw rate; (b) Sideslip angle.

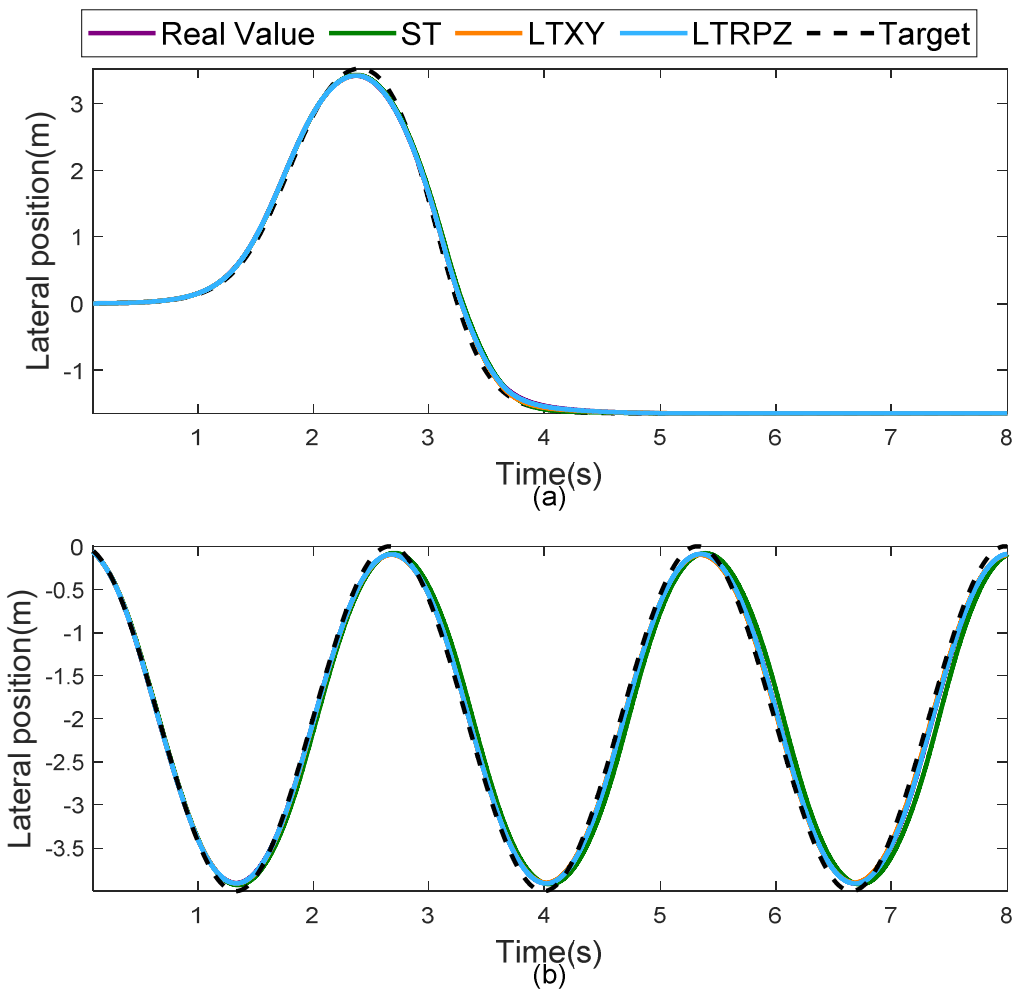

Fig. 10. Tracking performance with vertical load estimation. (a) Tracking path in DLC; (b) Tracking path in slalom.

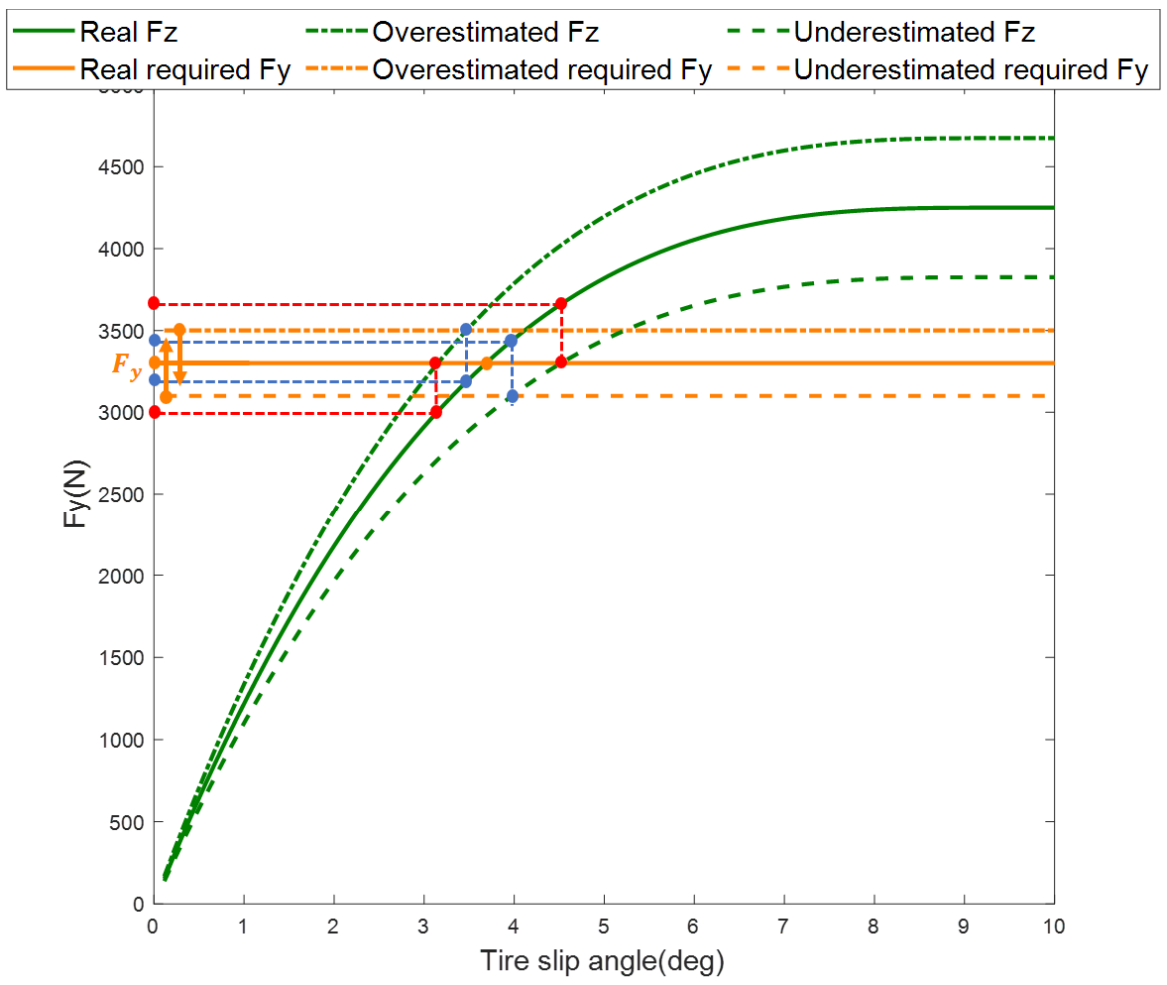

Fig. 11. Self-compensating effect with the actual realized lateral tire force. The red dots represent the lateral tire force from look-up table caused by load estimation errors. The blue dots indicate a self-compensation effect due to control allocation errors and look-up table errors, both resulting from load estimation inaccuracies.

In the control scheme of [9], the steering angle is determined directly from the motion controller. As the steering angle determination is irrelevant to the control allocation and actuator control layers, it is not influenced by the vertical load estimation. The vertical estimation primarily affects longitudinal force determination in the tire force allocation layer, leading to neglecting longitudinal and lateral coupling effects. To address this issue, we adopt a more integrated control scheme (Fig. 4), considering both longitudinal and lateral forces in the tire force allocation controller, as demonstrated in Section III. As discussed above, the overestimated vertical force and over-allocated lateral force result in a biased tire slip angle in the look-up table, as shown in Fig. 11. This, combined with the actual vertical load, results in the lateral force smaller than the over-allocated one, and with a minor difference from the real one as indicated by the blue dots. We call it a self-compensating effect, more like a negative feedback effect. The lateral force error exists, but is relatively small if the vertical load estimation bias is minimal, such as LTRPZ and LTXY.

### *B. Actuator Dynamic Characteristics*

Simulations are conducted with actuator dynamic characteristics incorporated into the model. In the step yaw rate input scenario, the control allocation method that considers actuator dynamic characteristics significantly improves the yaw rate response speed. This is comparable to the model without actuator dynamics, as Fig. 12 shows. Neglecting actuator dynamics in control leads to increased settling time due to response lag. Fig. 13 demonstrates that lateral tire force control, considering actuator dynamics, is faster by approximately 100 ms. Tire force tracking hysteresis results in the unexpected tracking error, as illustrated in Fig. 14. The tracking error for the method considering actuator dynamics is nearly 0, while the maximum error for the method neglecting actuator dynamics reaches 0.97 m. At handling limits, as tires usually work at their capacity, the unexpected tire force overshooting will saturate front or rear tires, making vehicles lose steering control or spin out, causing severe accidents. Fig. 15 shows that around 3.5 s, both the command and actual front tire forces in control allocation neglecting actuator dynamics exceed those considering actuator dynamics, causing front tire saturation and inability to track the ideal path during the DLC test. The path departure will panic drivers and potentially cause collisions with surrounding vehicles and facilities.

TABLE IV
ESTIMATION ERROR FOR VERTICAL LOAD IN THREE MANEUVERS

| Estimation Error | Step Yaw Rate Input | Double Lane Change | Slalom |
|---|---|---|---|
| ST | 89.35% | 55.24% | 100.68% |
| LTXY | 7.03% | 6.56% | 9.29% |
| LTRPZ | 2.85% | 4.29% | 5.16% |

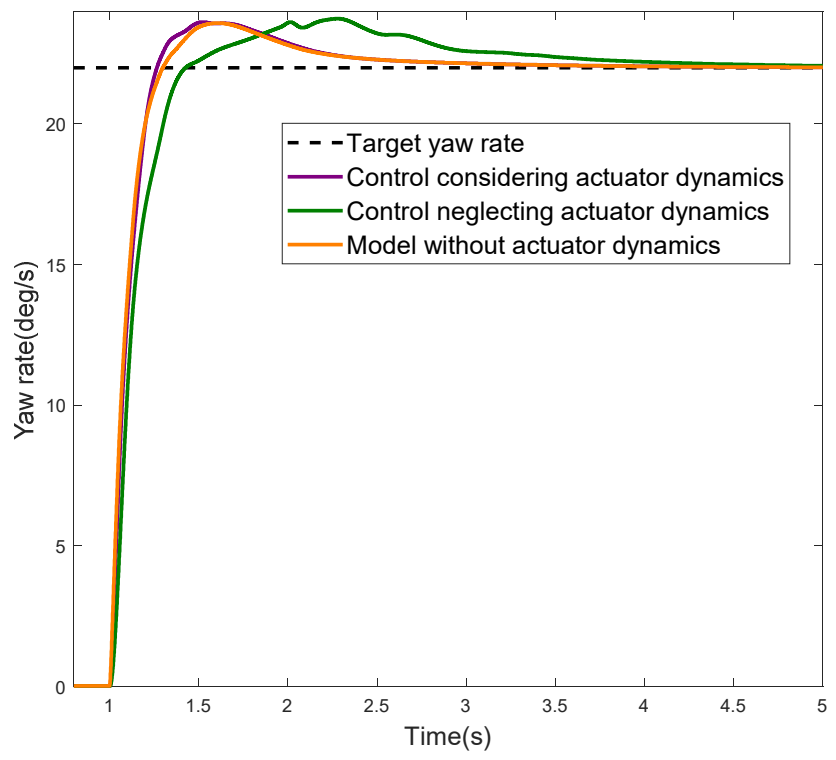

Fig. 12. Yaw rate response with actuator dynamic characteristics in step yaw rate input.

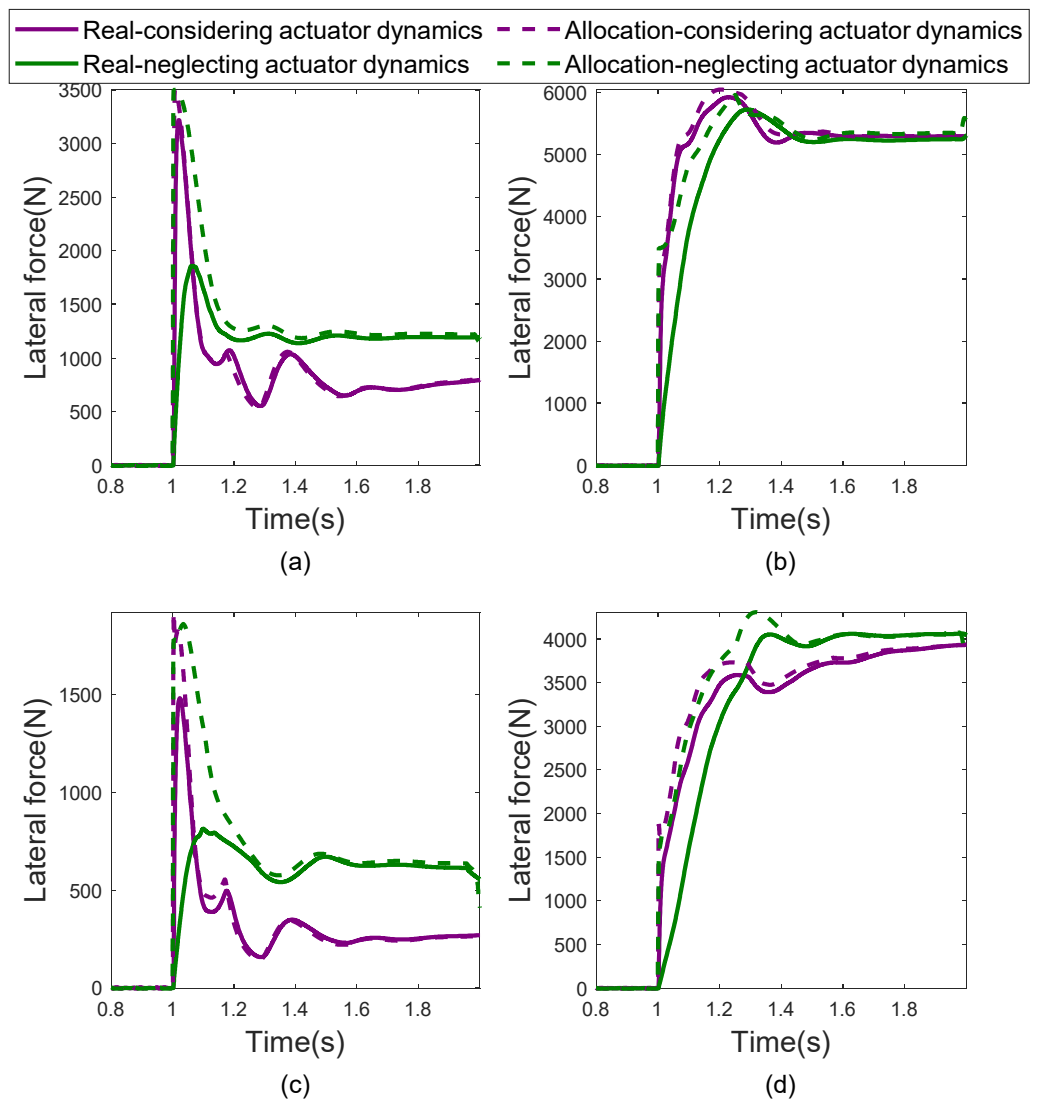

Fig. 13. Lateral force response with actuator dynamic characteristics in step yaw rate input. (a) Front left tire; (b) Front right tire; (c) Rear left tire; (d) Rear right tire.

In addition to response time, the discrepancy between commanded and actual tire forces is crucial. Ignoring actuator dynamics results in tire forces consistently lagging behind commands, increasing target control force demand through feedback control. This inertial lag causes actuators to consistently miss target control values, leading to a continuous decline in tracking performance, as depicted in Fig. 16. In the slalom test, the mean tracking error of the method considering actuator dynamics is 0.06 m, compared to 0.6 m for the method ignoring actuator dynamics.

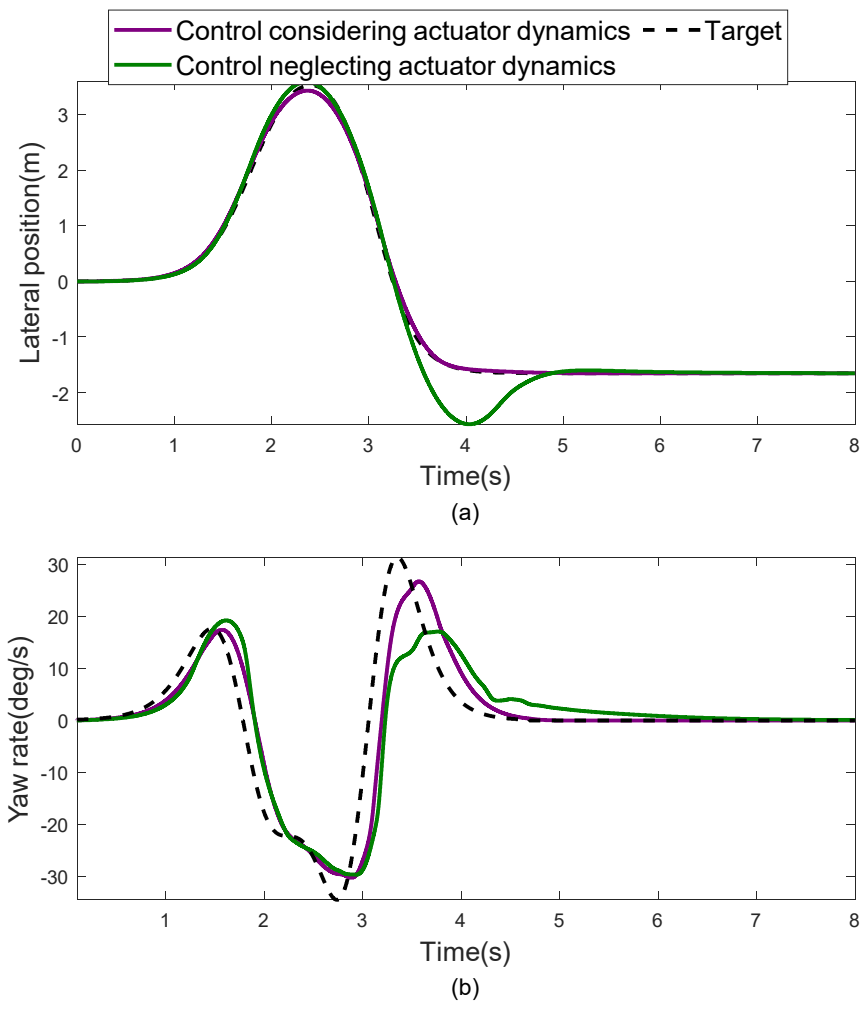

Fig. 14. Tracking performance with actuator dynamic characteristics in DLC. (a) Tracking path; (b) Yaw rate.

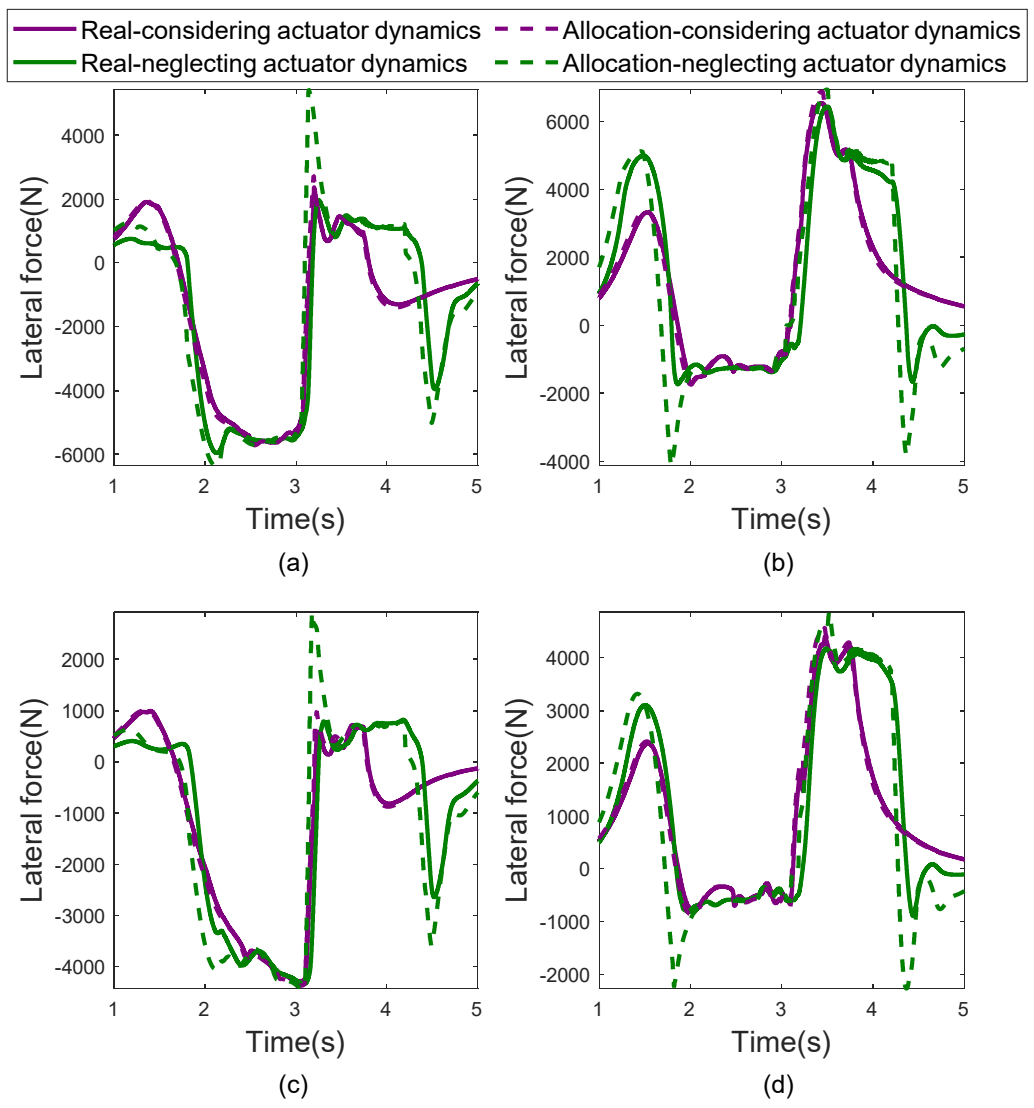

Fig. 15. Lateral force response with actuator dynamic characteristics in DLC. (a) Front left tire; (b) Front right tire; (c) Rear left tire; (d) Rear right tire.

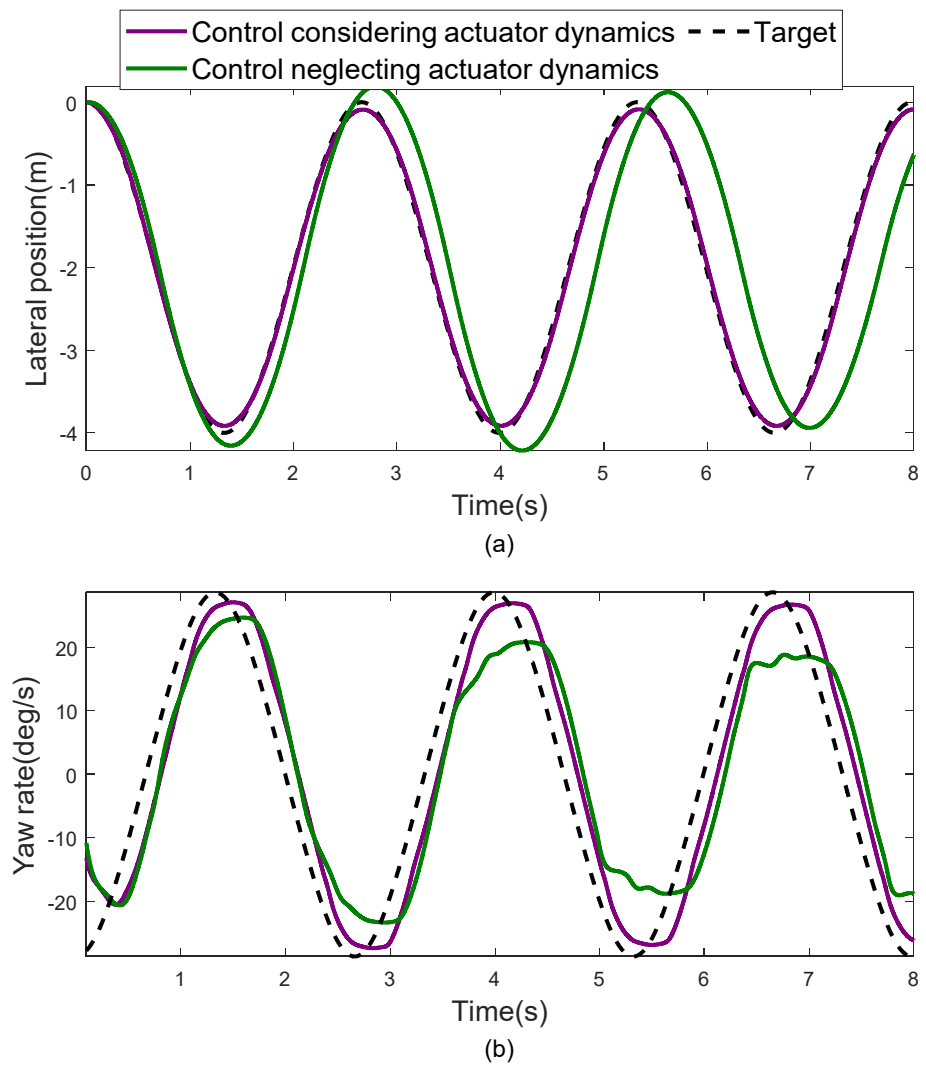

Fig. 16. Tracking performance with actuator dynamic characteristics in slalom. (a) Tracking path; (b) Yaw rate.

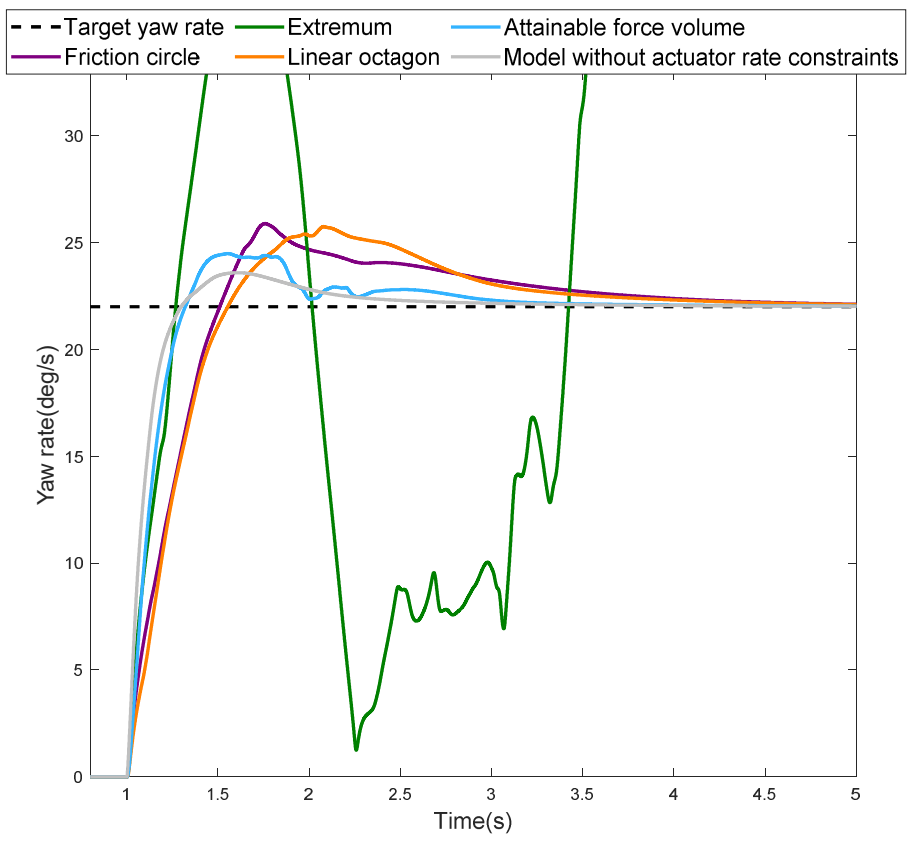

Fig. 17. Yaw rate response under different tire force constraints in step yaw rate input.

## C. Tire Force Constraints

The actuator rate constraints limit excessive changes in tire forces over short periods, even within the tire-road adhesion limits, particularly for steering actuators. In the step yaw rate input scenario, the yaw rate response slows down among all tire force constraints. However, the yaw rate response under the real-time attainable tire force volume, calculated with the proposed convex polygonal approximation method, closely matches the model without actuator rate constraints, as shown in Fig. 17. Under extremum constraints, a highly inaccurate estimation of the attainable tire force domain results in excessive yaw rate, leading to instability and lack of convergence. We use $Muy = f_{yi}/\mu f_{zi}$ and $Mux = f_{xi}/\mu f_{zi}$ to normalize the lateral and longitudinal friction utilization. Fig. 18 and Fig. 19 demonstrate the commanded and real tire forces from 1 s to 4 s.

Overestimation leads to unattainably allocated tire forces in reality, both in magnitude and direction, as depicted in Fig. 18. Under extremum constraints, allocated tire forces consistently reach the square boundary, ignoring the longitudinal and lateral coupling constraints of the tire force and actuator rate constraints. Under friction circle or linear octagon constraints, allocated tire forces are initially set at the constraint boundary, requiring significantly large lateral tire forces due to the step yaw rate input.

Under the attainable force volume, tire force demand changes moderately and is sensibly allocated. The real tire forces closely match the demand, as illustrated in Fig. 19. However, under the other three constraints, real physical actuator rate constraints cause tire forces to change gradually, resulting in a significant discrepancy from the allocated values. Misestimating tire force constraints complicates control of the actual tire forces, often leading to unexpected results. While tire force controller commands consistently reach constraint boundaries, real tire forces erratically contract within true limits due to physical constraints. This irregular contraction, stemming from inverse model error caused by a notable discrepancy between commanded and actual tire characteristics, highlights the importance of accounting for the real attainable domain of tire forces within road adhesion and actuator constraints.

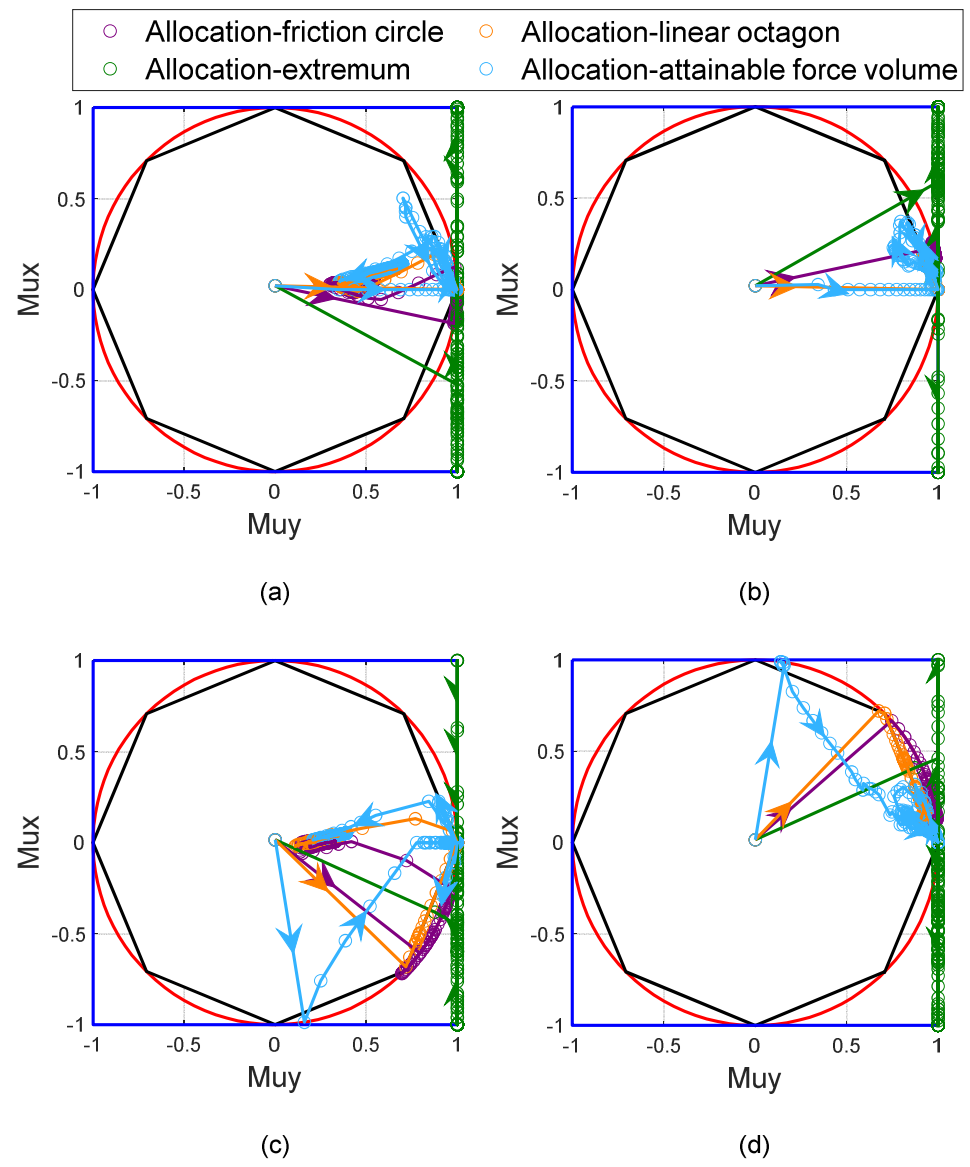

Fig. 18.The target tire forces allocated under different tire force constraints in step yaw rate input from 1 s to 4 s. The small dot is taken per 10 ms and the arrows indicate the sequence of time. (a) Front left tire; (b) Front right tire; (c) Rear left tire; (d) Rear right tire.

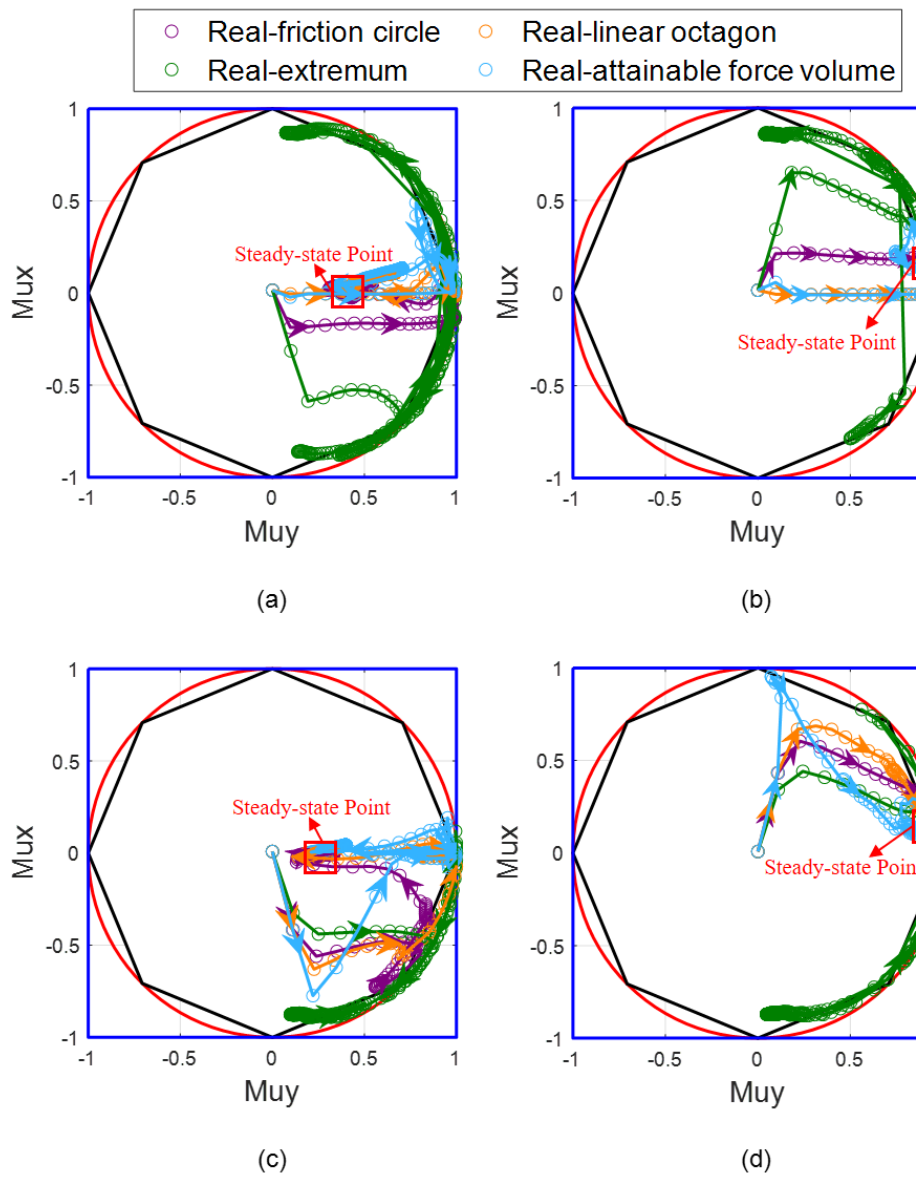

Fig. 19. The real tire forces under different tire force constraints in step yaw rate input from 1 s to 4 s. The small dot is taken per 10 ms and the arrows indicate the sequence of time. (a) Front left tire; (b) Front right tire; (c) Rear left tire; (d) Rear right tire.

In the DLC maneuver, ignoring certain constraints can lead to an overestimation of the attainable domain of tire forces, increasing the likelihood of unexpected tire forces and larger tracking errors. Tracking errors under extremum, friction circle, linear octagon constraints and attainable force volume are 2.12 m, 1.34 m, 0.50 m, and 0.31 m, respectively, as depicted in Fig. 20. Under extremum constraints, front tires become saturated, leading to push understeer and loss of trajectory tracking capability. Fig. 21 illustrates the target tire forces allocated and the real tire forces achieved under different tire force constraints in DLC at 3.1 s. It reveals that under friction circle and linear octagon constraints, tire force allocations at certain moments are unachievable due to ignored actuator rate constraints. This is particularly evident for the front left wheel, where the positive allocated lateral tire force contrasts with the opposite real force. Such misestimations can lead to vehicle instability during the control process, especially in extreme conditions, creating significant risks.

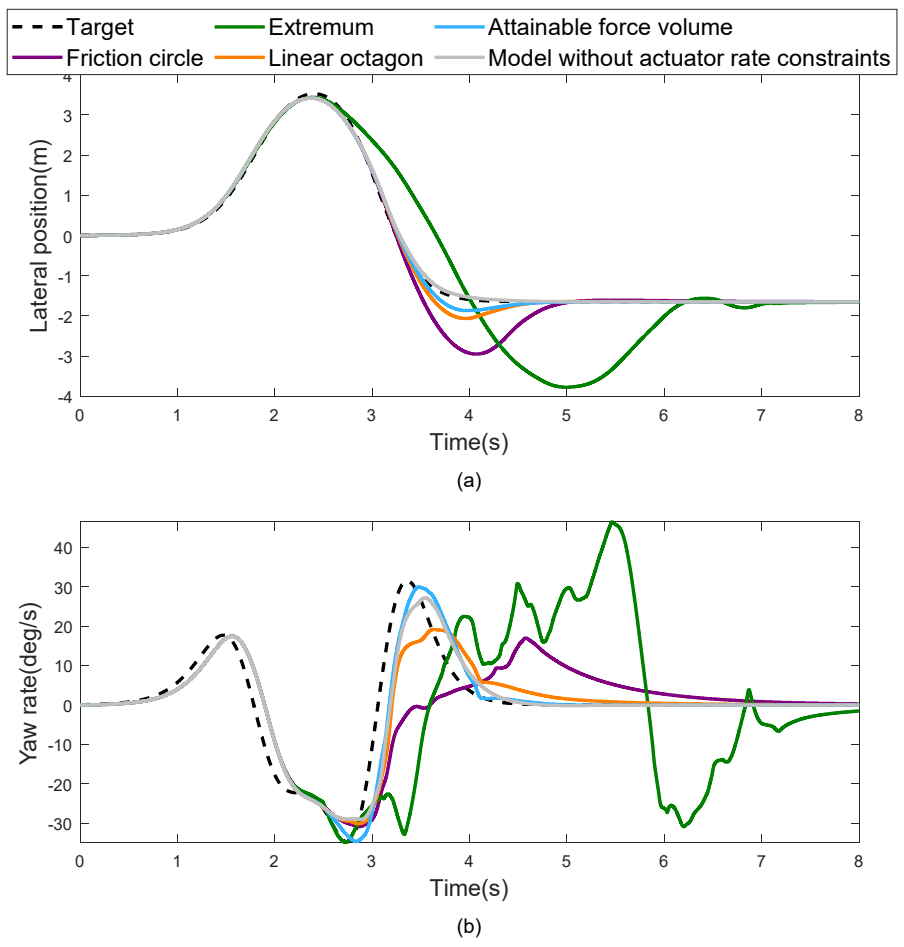

Fig. 20. Tracking performance under different tire force constraints in DLC. (a) Tracking path; (b) Yaw rate.

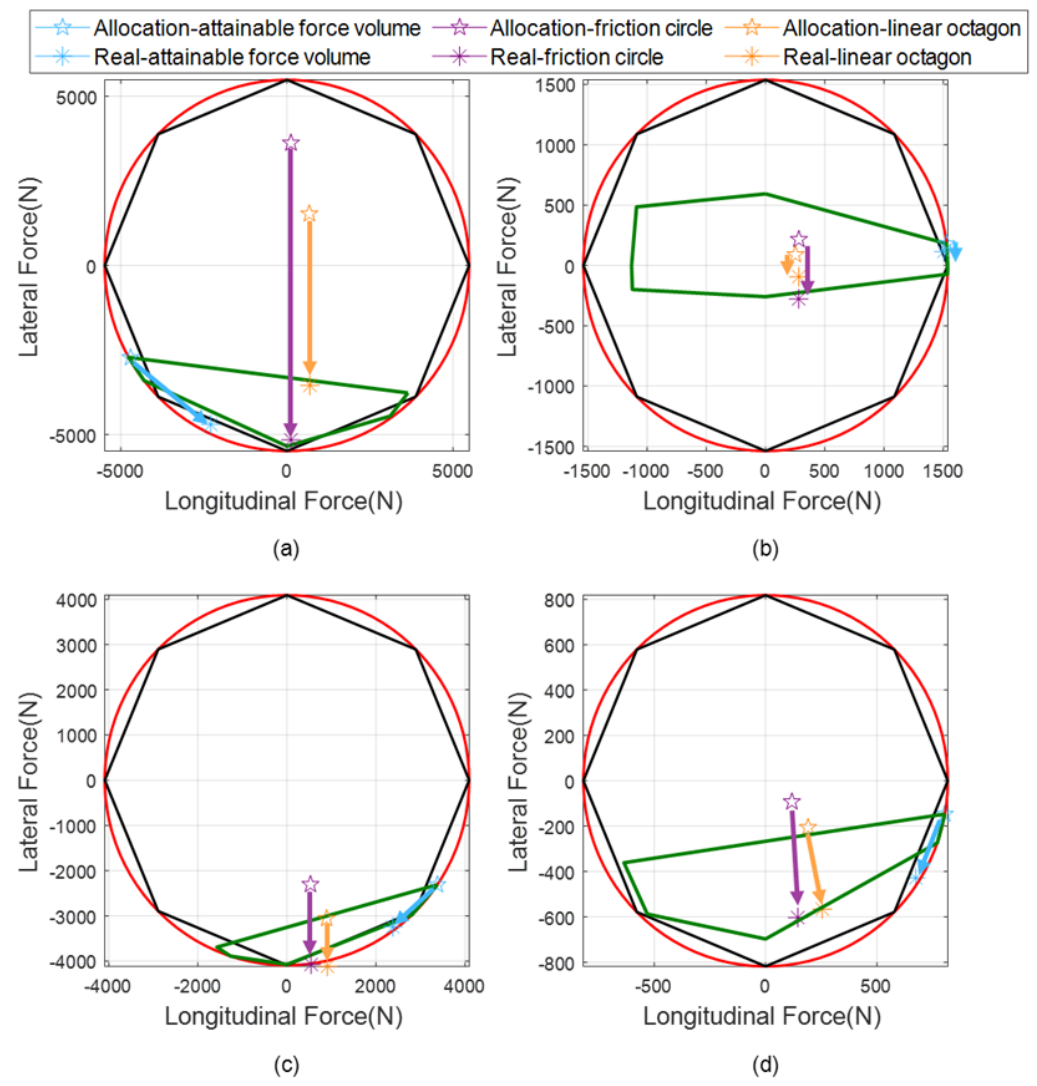

Fig. 21. The target tire forces allocated and the real tire forces achieved under different tire force constraints in DLC at 3.1 s. The green boundary represents the real-time attainable tire force volume calculated using the proposed convex polygonal approximation method. (a) Front left tire; (b) Front right tire; (c) Rear left tire; (d) Rear right tire.

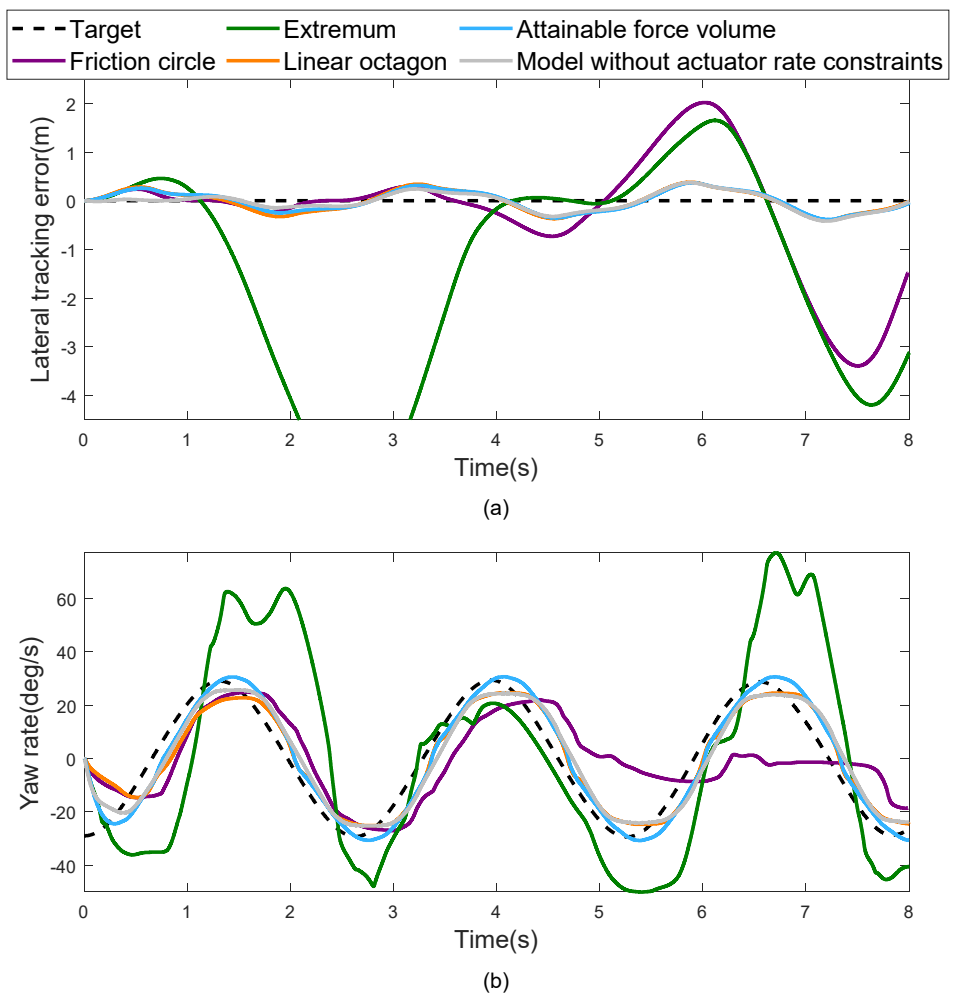

Fig. 22. Tracking performance under different tire force constraints in slalom. (a) Lateral tracking error; (b) Yaw rate.

In the slalom scenario, simulation results reconfirm the aforementioned explanation, as illustrated in Fig. 22. Larger deviations about the estimated attainable tire force domain lead to greater tracking errors. Control performance under attainable force volume aligns closely with the model without actuator rate constraints. The maximum trajectory tracking error under linear octagon constraints is slightly larger than that under attainable force volume, with a difference of 0.02 m.

### *D. Wheel Steering Precision*

Due to the proportional relationship between wheel toe angle and suspension jounce, the wheel toe angle increases as the outer suspension compresses and decreases while the inner suspension stretches during vehicle steering. This creates the bump steer that enlarges the steering angle for both left and right axles, causing them to steer in the same direction. Therefore, under the suspension kinematic characteristic curve, the lateral tire force generated by the normal steer is larger than the required value. In the step yaw rate input scenario, feedforward control with bump steer compensation results in a faster yaw rate response, as demonstrated in Fig. 23.

Under feedforward control, lateral tire force execution error is nearly zero at steady-state. In contrast, with normal steering, the average lateral force execution error of each tire is 192 N at steady-state, and the vehicle total lateral force execution error amounts to 752 N, as shown in Fig. 24. Lateral force error impacts tracking errors. Furthermore, ignoring bump steer leads to uniform deviation of all tire forces from actual values, causing either overuse or underuse of tire forces based on suspension characteristics. Additionally, significant fluctuations in tire force errors may lead to unanticipated tire force saturation, negatively affecting vehicle control performance at handling limits.

In both DLC and slalom maneuvers, no significant difference is observed between the two methods due to feedback adjustments made by the vehicle motion controller, as illustrated in Fig. 25. Although the impact of wheel steering precision differences on trajectory tracking is not marked under single-factor conditions, this does not imply the factor can be ignored. The importance of wheel steering precision is highlighted under the influence of multiple coupled factors, which will be detailed in the combined analysis.

### *E. Combined Analysis*

In this section, we examine the combined impacts of four key factors on control allocation for 4WID-4WIS electric vehicle dynamics, keeping controller parameters unchanged throughout. The analysis employs a baseline model incorporating actuator dynamic characteristics and rate constraints. The effects of actuator dynamics and tire force constraints are not the focus, due to their significant impacts in single-factor analysis. The comparison includes actuator dynamics in the control algorithm. The impacts of vertical load estimation and wheel steering precision are further explored. Four scenarios reflecting mixed influences of these key factors are examined: ① The optimal scenario with all factors integrated; ② Only omitting feedforward compensation from wheel steering angle control; ③ Replacing the vertical load estimation with LTXY; ④ Normal steering, employing LTXY for vertical load estimation, and adopting linear octagon constraints for tire force, which is the widely used approach. The simulation results are shown in Fig. 26 and Fig. 27.

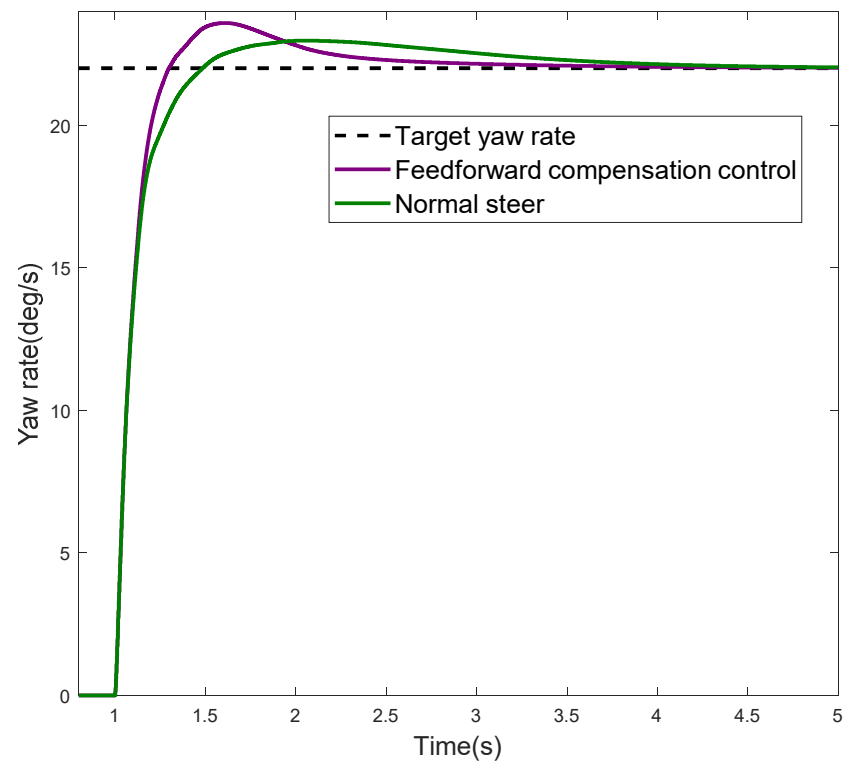

Fig. 23. Yaw rate response with wheel steering precision in step yaw rate input.

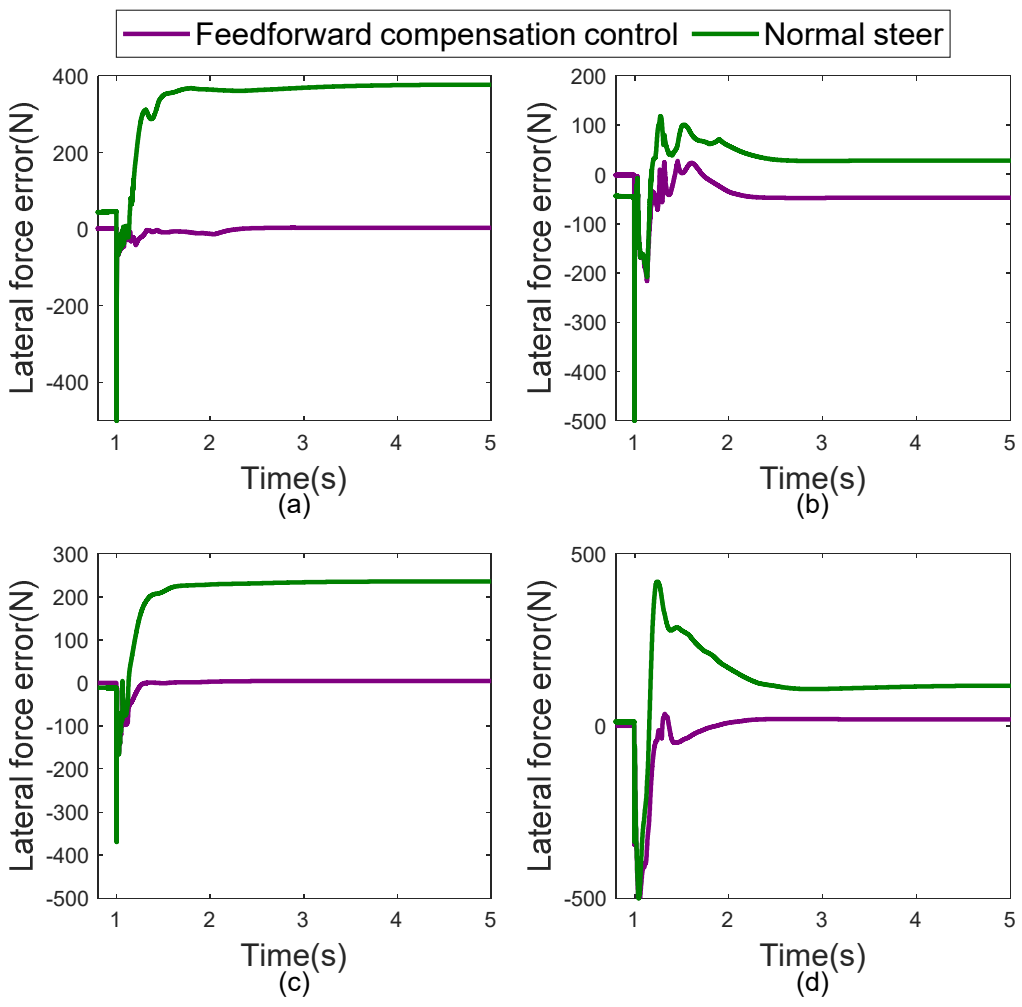

Fig. 24. Lateral tire force error in actual execution in step yaw rate input. (a) Front left tire; (b) Front right tire; (c) Rear left tire; (d) Rear right tire.

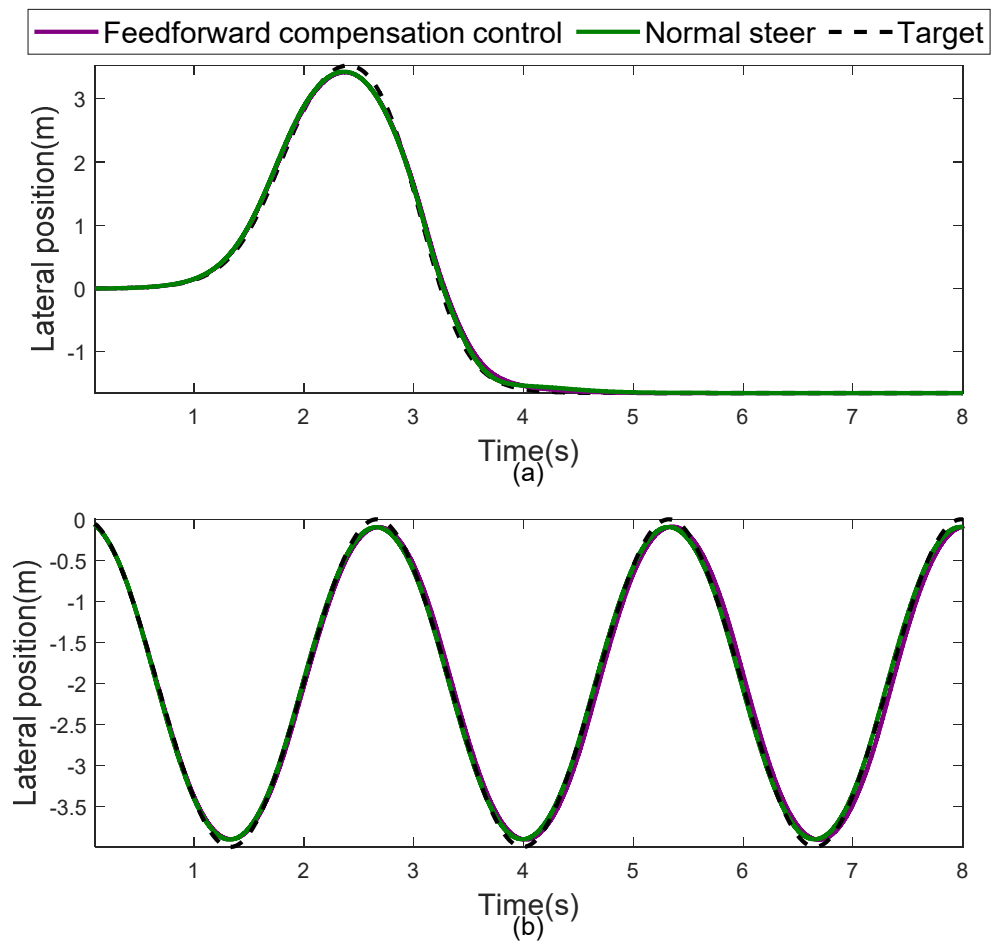

Fig. 25. Tracking performance with wheel steering precision. (a) Tracking path in DLC; (b) Tracking path in slalom.

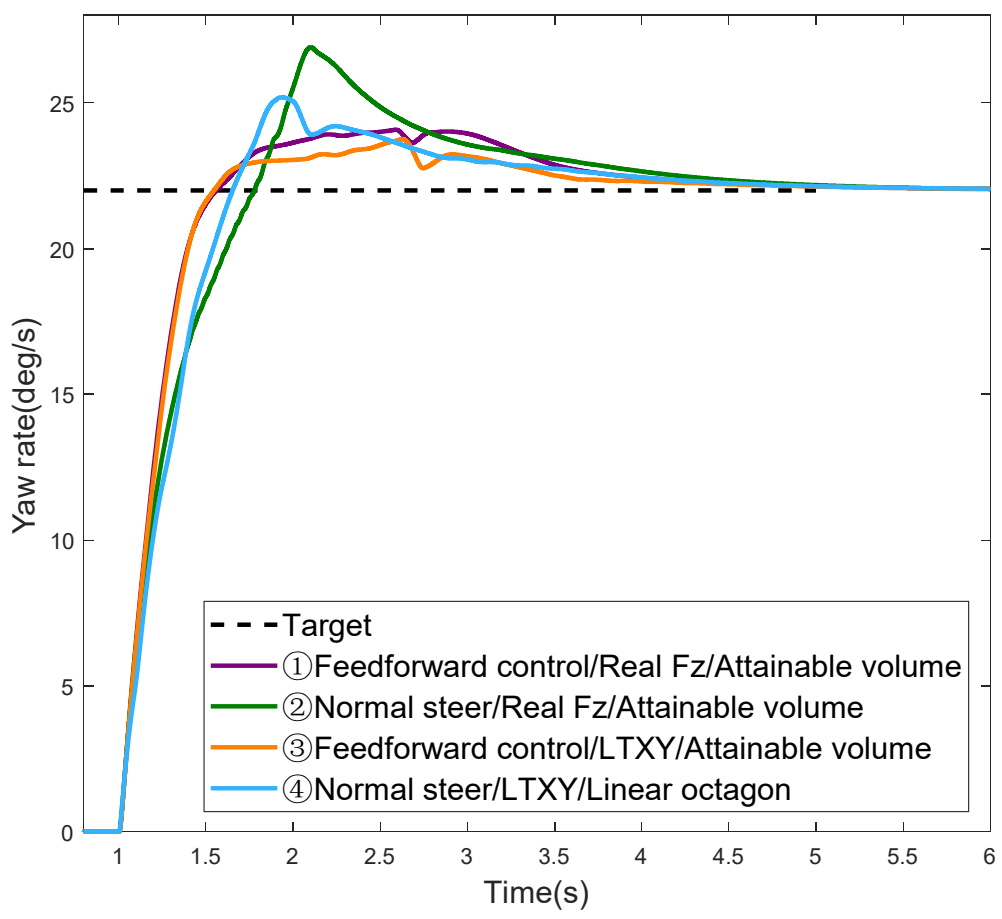

Fig. 26. Yaw rate response with combined analysis in step yaw rate input.

Under the combined influence of four factors, the impact of wheel steering precision becomes significant. Yaw rate response slows in the step yaw rate input scenario, and the trajectory tracking error notably increases in DLC, reaching up to 2.43 m. The attainable domain of tire forces changes in real-time, due to actuator rate constraints and tire-road adhesion limits. In such complex conditions, significant errors in lateral tire force control can easily lead to unanticipated tire force saturation, worsening vehicle control performance. However, when load estimation error is relatively small, its impact on yaw rate response and trajectory tracking is not significant, mainly due to the self-compensation effect of lateral tire forces in control allocation, as previously mentioned. This suggests that the load estimation, LTXY, is effective to some extent.

Compared to scenario ①, the yaw rate response under the widely used approach (scenario ④) slows in the step yaw rate input, and the maximum trajectory tracking error in DLC increases by 0.4 m. This further highlights the importance of considering these key factors during tire force control allocation.

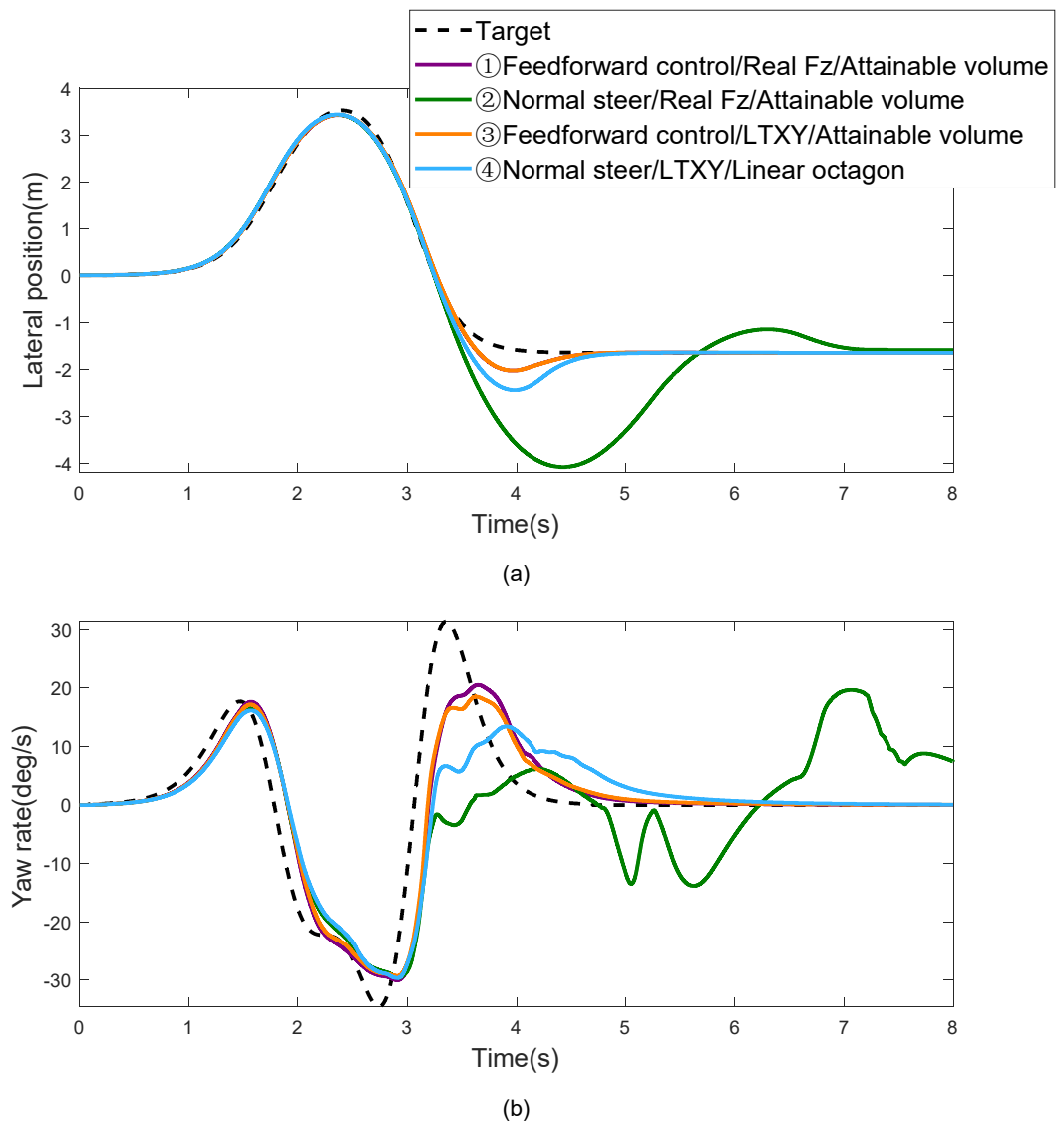

Fig. 27. Tracking performance with combined analysis in DLC. (a) Tracking path; (b) Yaw rate.

## V. Conclusion

This paper presents a hierarchical control allocation framework for more integrated electric vehicles with 4WID-4WIS configurations. It comprehensively examines the influences of four key factors, including vertical load estimation, actuator dynamic characteristics, tire force constraints, and wheel steering precision on tire force control allocation. Key findings are as follows:

1) Precise vertical load estimation enhances lateral force allocation accuracy, suppresses vehicle sideslip angle, and reduces response time. When the deviation in vertical load estimation is minimal, its impact on tire force control allocation becomes negligible due to the self-compensating effect of lateral tire forces.
2) The proposed control allocation method, considering actuator dynamic characteristics, significantly improves yaw rate response and minimizes tracking errors in various scenarios. Conversely, neglecting actuator dynamics can cause response lag, and tire force overshooting, which may result in losing steering control or spinning out.
3) The proposed convex polygonal method, based on $\alpha_i$ and $\kappa_i$, effectively approximates the real-time attainable force volume under tire-road adhesion and actuator constraints, thereby enhancing yaw rate response and trajectory tracking performance. Larger deviations about the estimated attainable domain of tire forces can lead to increased tracking errors.
4) Feedforward control, based on suspension kinematics, significantly improves wheel steering precision and lateral tire force control accuracy. Ignoring bump steer can result in either overuse or underuse of tire forces. The precision of tire force control notably affects yaw rate response and trajectory tracking in scenarios with combined factors.

These findings provide valuable insights for improving control allocation strategies in more integrated electric vehicles and optimizing vehicle dynamics control performance under various driving conditions.

## References

[1] X. Zhang, D. Göhlich, and J. Li, "Energy-efficient toque allocation design of traction and regenerative braking for distributed drive electric vehicles," *IEEE Trans. Veh. Technol.,* vol. 67, no. 1, pp. 285-295, 2017.

[2] L. Zhai, T. Sun, and J. Wang, "Electronic stability control based on motor driving and braking torque distribution for a four in-wheel motor drive

electric vehicle," *IEEE Trans. Veh. Technol.,* vol. 65, no. 6, pp. 4726-4739, 2016.
[3] X. Zhang, D. Göhlich, and W. Zheng, "Karush–Kuhn–Tuckert based global optimization algorithm design for solving stability torque allocation of distributed drive electric vehicles," *J. Franklin Inst.,* vol. 354, no. 18, pp. 8134-8155, 2017.
[4] J. Yoon, W. Cho, B. Koo, and K. Yi, "Unified chassis control for rollover prevention and lateral stability," *IEEE Trans. Veh. Technol.,* vol. 58, no. 2, pp. 596-609, 2008.
[5] N. Guo, B. Lenzo, X. Zhang, Y. Zou, R. Zhai, and T. Zhang, "A real-time nonlinear model predictive controller for yaw motion optimization of distributed drive electric vehicles," *IEEE Trans. Veh. Technol.,* vol. 69, no. 5, pp. 4935-4946, 2020.
[6] H. Park and J. C. Gerdes, "Optimal tire force allocation for trajectory tracking with an over-actuated vehicle," in *IEEE Intell. Veh. Symp. Proc.*, Seoul, Korea, 2015, pp. 1032-1037.
[7] Y. Tao, X. Xie, H. Zhao, W. Xu, and H. Chen, "A regenerative braking system for electric vehicle with four in-wheel motors based on fuzzy control," in *Chinese Control Conf. (CCC)*, Dalian, China, Jul. 26-28, 2017, pp. 4288-4293.
[8] J. Talbot, M. Brown, and J. C. Gerdes, “Shared Control up to the Limits of Vehicle Handling,” IEEE Trans. Intell. Veh., pp. 1–11, 2023.
[9] R. Li, Y. Sun, Z. Lu, and G. Tian, "Tire Force Allocation with Different Vertical Load Estimation Methods for 4WID-4WIS Vehicles," in *IEEE Veh. Power Propuls. Conf. (VPPC)*, Merced, CA, Nov. 1-4, 2022, pp. 1-6.
[10] A. Rezaeian *et al.*, "Novel Tire Force Estimation Strategy for Real-Time Implementation on Vehicle Applications," *IEEE Trans. Veh. Technol.,* vol. 64, no. 6, pp. 2231-2241, 2015.
[11] W. Cho, J. Yoon, S. Yim, B. Koo, and K. Yi, "Estimation of tire forces for application to vehicle stability control," *IEEE Trans. Veh. Technol.,* vol. 59, no. 2, pp. 638-649, 2009.
[12] W. Zhang, Z. Wang, L. Drugge, and M. Nybacka, "Evaluating Model Predictive Path Following and Yaw Stability Controllers for Over-Actuated Autonomous Electric Vehicles," *IEEE Trans. Veh. Technol.,* vol. 69, no. 11, pp. 12807-12821, 2020.
[13] L. Zhai, T. Sun, and J. Wang, “Electronic Stability Control Based on Motor Driving and Braking Torque Distribution for a Four In-Wheel Motor Drive Electric Vehicle,” *IEEE Trans. Veh. Technol.*, vol. 65, no. 6, pp. 4726–4739, 2016.
[14] Z. Shuai, H. Zhang, J. Wang, J. Li, and M. Ouyang, “Lateral motion control for four-wheel-independent-drive electric vehicles using optimal torque allocation and dynamic message priority scheduling,” *Control Eng. Pract.*, vol. 24, pp. 55–66, 2014.
[15] Y. Chen and J. Wang, "Design and Experimental Evaluations on Energy Efficient Control Allocation Methods for Overactuated Electric Vehicles: Longitudinal Motion Case," *IEEE/ASME Trans. Mechatron.,* vol. 19, no. 2, pp. 538-548, 2014.
[16] M. Hanger, T. A. Johansen, G. K. Mykland, and A. Skullestad, "Dynamic model predictive control allocation using CVXGEN," in *IEEE Int. Conf. Control Autom. (ICCA)*, Santiago, Chile, Dec. 19-21, 2011, pp. 417-422.
[17] A. Wong, D. Kasinathan, A. Khajepour, S.-K. Chen, and B. Litkouhi, "Integrated torque vectoring and power management framework for electric vehicles," *Control Eng. Pract.,* vol. 48, pp. 22-36, 2016.
[18] J. Guo, Y. Luo, and K. Li, “An adaptive hierarchical trajectory following control approach of autonomous four-wheel independent drive electric vehicles,” *IEEE Trans. Intell. Transp. Syst.*, vol. 19, no. 8, pp. 2482-2492, 2017.
[19] M. Jonasson and J. Andreasson, "Exploiting autonomous corner modules to resolve force constraints in the tyre contact patch," *Veh. Syst. Dyn.,* vol. 46, no. 7, pp. 553-573, 2008.
[20] H. Zhao, W. Chen, J. Zhao, Y. Zhang, and H. Chen, "Modular integrated longitudinal, lateral, and vertical vehicle stability control for distributed electric vehicles," *IEEE Trans. Veh. Technol.,* vol. 68, no. 2, pp. 1327-1338, 2019.
[21] R. de Castro, M. Tanelli, R. E. Araújo, and S. M. Savaresi, "Design of safety-oriented control allocation strategies for overactuated electric vehicles," *Veh. Syst. Dyn.,* vol. 52, no. 8, pp. 1017-1046, 2014.
[22] Y. Wang, W. Deng, B. Zhu, Q. Zhao, and B. Litkouhi, "Allocation-Based Control with Actuator Dynamics for Four-Wheel Independently Actuated Electric Vehicles," *SAE Int. J. Passeng. Cars - Electron. Electr. Syst.,* vol. 8, no. 2015-01-0653, pp. 425-432, 2015.
[23] X. Sun, Z. Shi, Y. Cai, G. Lei, Y. Guo, and J. Zhu, “Driving-Cycle-Oriented Design Optimization of a Permanent Magnet Hub Motor Drive System for a Four-Wheel-Drive Electric Vehicle,” *IEEE Trans. Transp. Electrific.*, vol. 6, no. 3, pp. 1115–1125, 2020.
[24] G. Mohan, C. Rouelle, and E. Hugon, "A new method to evaluate bump steer and steering influence on kinematic roll and pitch axes for all independent suspension types," SAE Tech. Paper, 0148-7191, 2008.
[25] U. Kulkarni, M. M. Gowda, and H. K. Venna, "Effect of Tie Rod Length Variation on Bump Steer," SAE Tech. Paper, 0148-7191, 2016.
[26] A. Lidozzi, L. Solero, F. Crescimbini, and A. Di Napoli, "Direct tuning strategy for speed controlled PMSM drives," in *IEEE Int. Symp. Ind. Electron.*, 2010, pp. 1265-1270.
[27] C. Satzger and R. de Castro, "Predictive brake control for electric vehicles," *IEEE Trans. Veh. Technol.,* vol. 67, no. 2, pp. 977-990, 2017.
[28] R. Hou, L. Zhai, T. Sun, Y. Hou, and G. Hu, "Steering stability control of a four in-wheel motor drive electric vehicle on a road with varying adhesion coefficient," *IEEE Access,* vol. 7, pp. 32617-32627, 2019.
[29] Y. Wang, " Studies on Integrated Vehicle Controls based on Constrained Optimization," Ph.D. dissertation, Dept. Auto. Eng., Jilin Univ., Jilin, China, 2016.
[30] G. Rill, "First order tire dynamics," in *Proc. 3rd European Conference on Computational Mechanics Solids, Structures and Coupled Problems in Engineering*, Lisbon, Portugal, 2006, vol. 58.
[31] P. Dai and J. Katupitiya, “Force control for path following of a 4WS4WD vehicle by the integration of PSO and SMC,” *Veh. Syst. Dyn.*, vol. 56, no. 11, pp. 1682–1716, 2018.
[32] R. Li, Y. Yu, Y. Sun, Z. Lu, and G. Tian, "Trajectory Following Control for Automated Drifting of 4WID Vehicles," SAE Tech. Paper, 0148-7191, 2022.
[33] V. Fors, B. Olofsson, and L. Nielsen, “Attainable force volumes of optimal autonomous at-the-limit vehicle manoeuvres,” *Veh. Syst. Dyn.*, vol. 58, no. 7, pp. 1101–1122, 2020.
[34] E. Ono, Y. Hattori, Y. Muragishi, and K. Koibuchi, “Vehicle dynamics integrated control for four-wheel-distributed steering and four-wheel-distributed traction/braking systems,” *Veh. Syst. Dyn.*, vol. 44, no. 2, pp. 139–151, 2006.
[35] Controllability and stability test procedure for automobile, GB/T 6323-2014.
[36] O. Mokhiamar and M. Abe, "Simultaneous optimal distribution of lateral and longitudinal tire forces for the model following control," *J. Dyn. Sys., Meas., Control,* vol. 126, no. 4, pp. 753-763, 2004.